\documentclass{ieeeaccess}
\usepackage{cite}
\usepackage{amsmath,amssymb,amsfonts}
\usepackage{algorithmic}
\usepackage{graphicx}
\usepackage{textcomp}

\usepackage{amsmath,mathtools,bm,amssymb}
\usepackage{enumitem}
\usepackage{xspace}
\usepackage{pifont}
\usepackage{dsfont}
\usepackage{graphicx}
\usepackage{subfig}
\usepackage{grffile}
\usepackage{pdflscape}
\usepackage{breqn}
\usepackage{amsthm}
\usepackage{bm}
\usepackage{color}
\usepackage{braket}

\renewcommand\bra[1]{{\langle{#1}|}}
\makeatletter
\renewcommand\ket[1]{%
  \@ifnextchar\bra{\k@t{#1}\!}{\k@t{#1}}%
}
\newcommand\k@t[1]{{|{#1}\rangle}}
\makeatother

\allowdisplaybreaks
\graphicspath{{./figs/}}

\newcommand{\vect}[1]{\boldsymbol{#1}}

\newcommand{\eg}{\textit{e.g.}}
\newcommand{\ie}{\textit{i.e.}}

\def\BibTeX{{\rm B\kern-.05em{\sc i\kern-.025em b}\kern-.08em
    T\kern-.1667em\lower.7ex\hbox{E}\kern-.125emX}}
\begin{document}
\doi{}

\title{Quantum Network Utility Maximization}
\author{\uppercase{Gayane Vardoyan\authorrefmark{1,3}}, 
\uppercase{Stephanie Wehner}\authorrefmark{1,2,3}}
\address[1]{QuTech, Delft University of Technology}
\address[2]{Kavli Institute of Nanoscience, Delft University of Technology}
\address[3]{Quantum Computer Science, Electrical Engineering, Mathematics and Computer Science, Delft University of Technology}
\address{emails: g.s.vardoyan@tudelft.nl, s.d.c.wehner@tudelft.nl}
\if{false}
\markboth
{Author \headeretal: Preparation of Papers for IEEE Transactions on Quantum Engineering}
{Author \headeretal: Preparation of Papers for IEEE Transactions on Quantum Engineering}

\corresp{Corresponding author: First A. Author (email: author@ boulder.nist.gov).}
\fi
\begin{abstract}
Network Utility Maximization (NUM) is a mathematical framework that has endowed researchers with powerful methods for designing and analyzing classical communication protocols. NUM has also enabled the development of distributed algorithms for solving the resource allocation problem, while at the same time providing certain guarantees, \eg, that of fair treatment, to the users of a network. We extend here the notion of NUM to quantum networks, and propose three quantum utility functions -- each incorporating a different entanglement measure. We aim both to gain an understanding of some of the ways in which quantum users may perceive utility, as well as to explore structured and theoretically-motivated methods of simultaneously servicing multiple users in distributed quantum systems. Using our quantum NUM constructions, we develop an optimization framework for networks that use the single-photon scheme for entanglement generation, which enables us to solve the resource allocation problem while exploring rate-fidelity tradeoffs within the network topologies that we consider. We learn that two of our utility functions, which are based on distillable entanglement and secret key fraction, are in close agreement with each other and produce similar solutions to the optimization problems we study. Our third utility, based on entanglement negativity, has more favorable mathematical properties, and tends to place a higher value on the rate at which users receive entangled resources, compared to the two previous utilities, which put a higher emphasis on end-to-end fidelity. These contrasting behaviors thus provide ideas regarding the suitability of quantum network utility definitions to different quantum applications.
\end{abstract}


\titlepgskip=-15pt

\maketitle

\section{Introduction}
\label{sec:introduction}
Quantum networks enable a host of applications whose benefits are impossible to glean with classical means alone. Notable examples of such capabilities are provably secure communication \cite{bennet1984quantum,ekert1992quantum}, improved sensing \cite{giovannetti2004quantum,jozsa2000quantum}, and blind quantum computation (BQC) \cite{broadbent2009universal,fitzsimons2017unconditionally}. To support such applications, quantum networks are expected to be able to produce and distribute entangled states of sufficiently high quality to nodes that request them. Efficient entanglement distribution is thus a task that is central to quantum communication, and one that is especially consequential during the noisy intermediate-scale quantum (NISQ) era \cite{preskill2018quantum}. Further, settings where multiple user groups demand quantum resources from a network pose a challenge not only in determining optimal entanglement distribution algorithms, but also in the very definition and interpretation of optimality -- whether from the perspective of a single user group or from that of the entire network. 

To address the need for quantifying a network's ability to serve user needs, we present here the notion of \textit{quantum network utility}, which serves as an analogue to the classical concept of network utility. The latter framework was first proposed in the seminal work of Kelly \textit{et al.} \cite{kelly1997charging,kelly1998rate} as a means of defining utility in classical communication networks, and subsequently being able to optimally (or fairly) allocate resources so as to maximize the sum of user or application utilities -- a process that is known as Network Utility Maximization (NUM). NUM's introduction to the networking community spurred a large body of research that resulted in both a rich theory as well as a family of algorithms for distributed resource allocation, some of which were even commercialized \cite{jin2004fast}. 
\begin{figure*}[ht]
\centering
\subfloat{\includegraphics[width=0.4\linewidth]{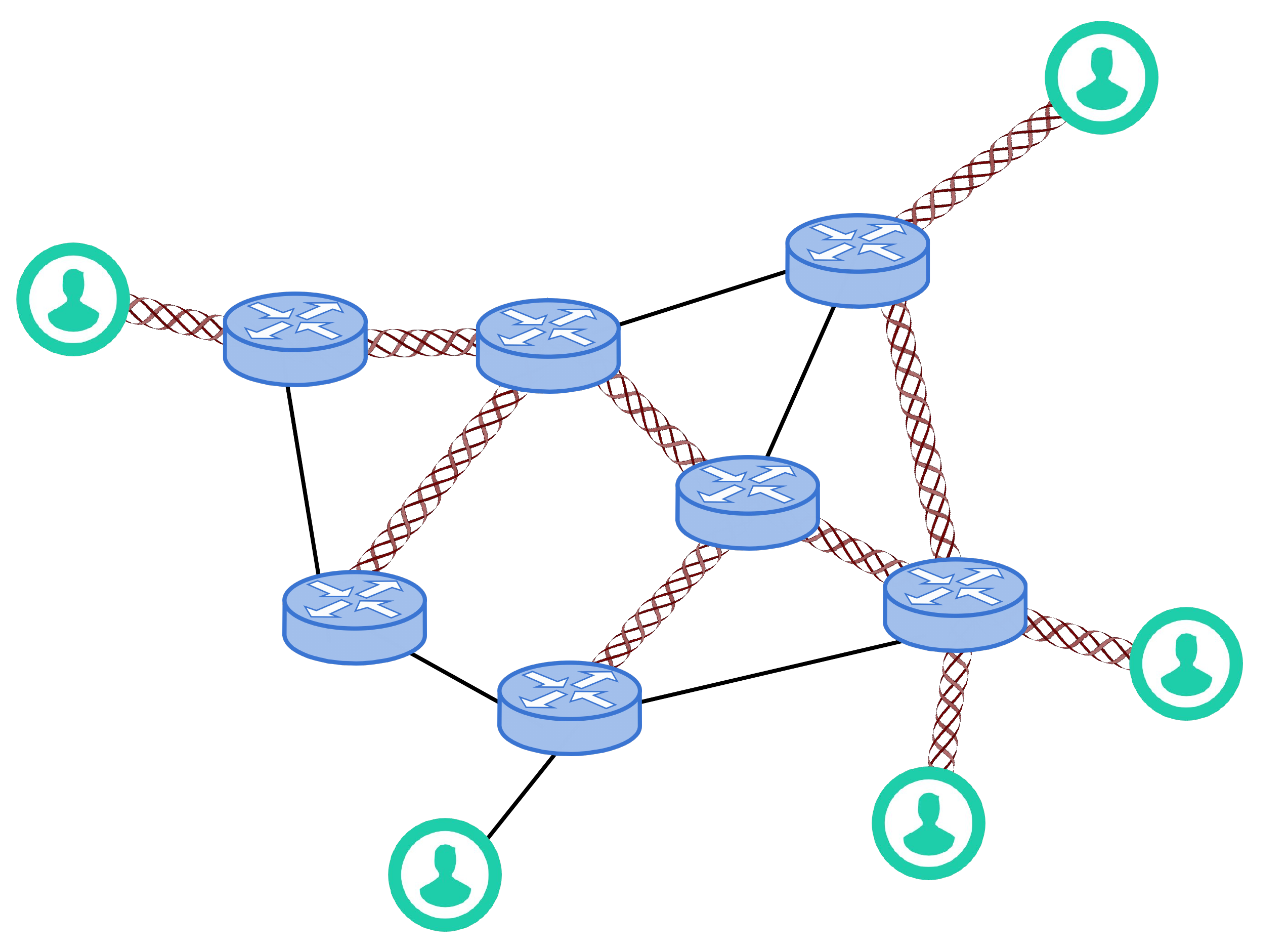}}
\hspace{-2.5cm}
\subfloat{\includegraphics[width=0.4\textwidth]{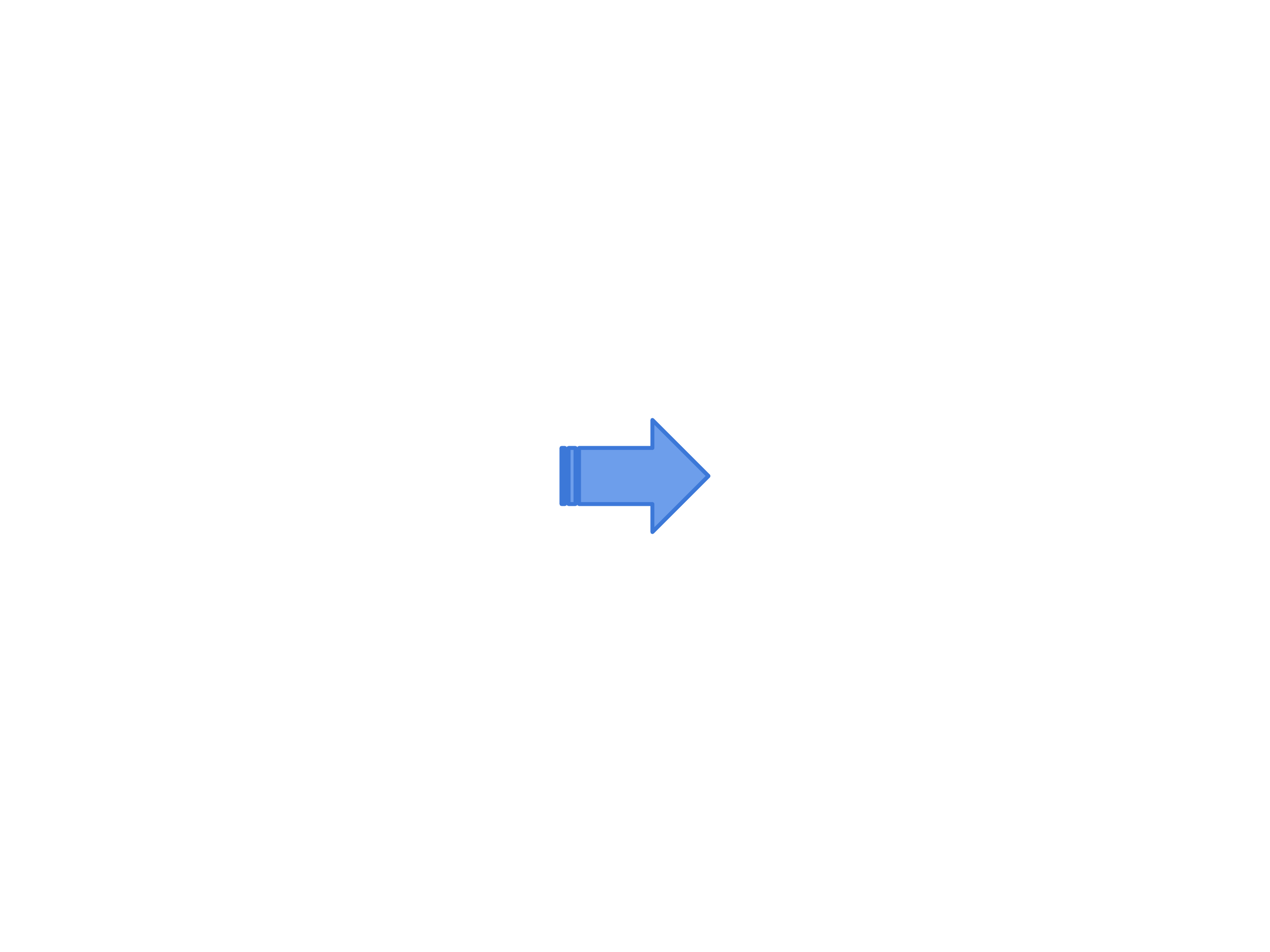}}
\hspace{-2.5cm}
\subfloat{\includegraphics[width=0.4\linewidth]{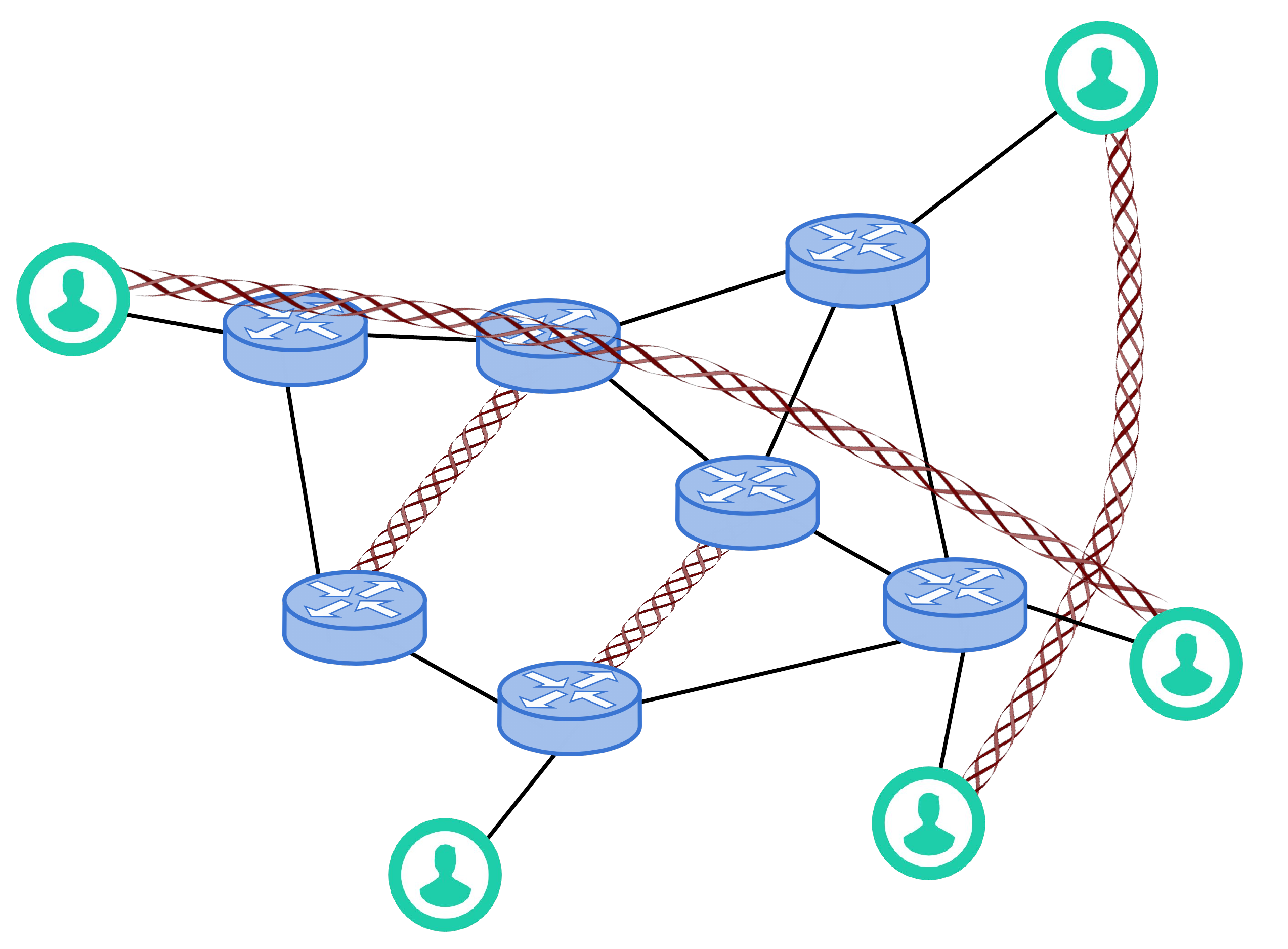}}
\caption{A quantum network consisting of a number of users and quantum repeaters. After links successfully generate entanglement (left), repeaters perform entanglement swapping to distribute end-to-end states to users (right).}
\label{fig:qnetEx}
\end{figure*}

By adapting the classical NUM framework to quantum networks, we aim to enable the quantum networking community to model and analyze distributed quantum architectures within a formal mathematical framework, potentially leading to the design of novel distributed resource allocation protocols that can service the needs of all communicating parties in a network. In extending the notion of NUM to quantum systems, however, one is inevitably confronted with a number of uniquely quantum problems:
\begin{itemize}
\item[(1)] \textit{Which methods can be used to cope with resource interdependence in quantum networks?}
In classical networks, physical channels serve as conduits through which data flows from a set of sources to a set of destinations. In contrast to the classical case, however, there is no notion of ``traffic'' in networks that distribute entanglement by first generating ``elementary'' entangled links between neighboring nodes and later converting these into longer-distance links via entanglement swapping \cite{zukowski1993event,pan1998experimental} (\ie, quantum networks made up of first- and second-generation quantum repeaters \cite{munro2015inside}), see Figure~\ref{fig:qnetEx} for an illustration. The network instead consists of independently-operating physical links that probabilistically produce resources -- entanglement -- which then, after undergoing a collective transformation, are consumed by groups of users. These state transformations  (\eg, entanglement swapping, distillation) result in interdependence between the nodes, links, and the states that they generate.
\item[(2)] \textit{How does a quantum user (an end node that runs a quantum application) derive utility from the network?} As with classical network utility, the precise answer to this question has a dependence on the application under consideration. In a classical network, an individual user's utility may be expressed in terms of rate, delay, jitter, and other perceptions of quality of service (QoS). An e-mail sender, for example has different requirements and expectations than a video watcher. 
In the same way, users who, for example, execute quantum key distribution (QKD) have a different understanding of QoS than a user who runs BQC.
However, in a quantum network, since entanglement is consumed by groups of users, an individual user's utility is always coupled to at least one other user's utility, and this joint utility in turn has a dependence both on the rate at which entanglement is allotted, as well as on the quality of each state. A utility function should then capture both classical (\eg, rate) as well as quantum (\eg, fidelity) QoS measures. 
\item[(3)] \textit{How should rate-fidelity tradeoffs, which are inherent in quantum networks, be captured?} In a classical network, a source may suffer diminishing returns as its sending rate is increased beyond a threshold, but it does not typically experience a \textit{decrease} in utility; \ie, traditionally, utility functions are concave and non-decreasing. In a quantum network, in contrast, one pays a cost for generating states with higher fidelity -- whether this is done via entanglement distillation or through other means (\eg, by adjusting the bright-state population when performing the single-photon entanglement generation protocol \cite{cabrillo1999creation,humphreys2018deterministic}) -- a decrease in the rate at which such states are successfully generated. A consequence of this is that utility is not necessarily a non-decreasing function of classical \textit{or} quantum QoS measures, and may in fact approach $0$ (or $-\infty$, depending on the range of the utility function) in cases where performance favors one measure exceedingly more than the others.
\end{itemize}

We make the following contributions towards answering these questions:
\begin{itemize}
\item We define and study three quantum network utility functions, each based on a different entanglement measure: distillable entanglement \cite{bennett1996mixed,rains2001semidefinite}, secret key rate \cite{shor2000simple,lo2005efficient}, and entanglement negativity \cite{vidal2002computable};
\item We develop a quantum NUM (QNUM) framework based on the single-photon entanglement generation scheme;
\item We apply the framework to a number of network topologies and solve the resource allocation problem by optimizing the rate-fidelity tradeoff on each physical link in the network.
\end{itemize}
We focus on applications that require bipartite entanglement generation between end users, to which we also refer to as end-to-end bipartite entanglement, as this is the most basic form of entanglement that is nevertheless a vital component of a diverse set of distributed quantum applications. To capture rate-fidelity tradeoffs as discussed in (3) above, we use the single-photon entanglement generation scheme, wherein the bright-state population parameter directly affects both the quality of states and the rate at which they are generated (see more in Sections \ref{sec:relatedwork} and \ref{sec:probform}). Entanglement distillation, while a necessary process for long-distance entanglement distribution in near-term networks, is left as a subject of follow-up work.

The remainder of this manuscript is organized as follows: in Section \ref{sec:relatedwork}, we provide relevant background on classical NUM. In Section \ref{sec:probform}, we state our assumptions and define the problem in detail. In Section \ref{sec:qnum}, we define three QNUM formulations and apply them to different network scenarios in Section \ref{sec:numexamples}. We conclude in Section \ref{sec:concl}.

\section{Background}
\label{sec:relatedwork}
NUM in its canonical form is the optimization problem
\begin{align}
\max\limits_{ x_r} \sum\limits_{r\in\mathcal{R}}U_r(x_r),
\end{align}
subject to
\begin{align}
\sum\limits_{r:l\in r}x_r&\leq c_l,~\forall l\in\mathcal{L},\\
x_r&\geq 0, ~\forall r\in\mathcal{R},
\end{align}
where $x_r$ corresponds to the rate at which a user sends packets across route $r$, $c_l$ is link $l$'s capacity, and $U_r$ is a utility function \cite{srikant2013communication}. $U_r$ is usually assumed to be strictly concave and may be used in one of two ways: either to describe the way in which a user (typically a source sending data) derives satisfaction from the network, or as a function that is \textit{assigned} to a user by the network to control the manner in which resources are allocated. A popular example of the latter is the idea of fair treatment, specifically, functions of the form
\begin{align}
U_r(x_r) = \frac{x_r^{1-\alpha}}{1-\alpha},
\label{eq:alpha_fair}
\end{align}
for $\alpha>0$, are known as $\alpha$-fair, each enforcing a different notion of utility. In this work, we will make use of the so-called proportionally-fair utility function $\log(x_r)$, which is obtained by letting $\alpha \to 1$ in (\ref{eq:alpha_fair}).

The classical NUM formulation stated above can be used as a starting point for developing distributed resource allocation algorithms which can operate in the presence of feedback delays and changing network conditions.
Following its development in the late nineties, NUM has proven itself highly influential in the design of distributed classical network algorithms (\eg, CSMA scheduling for multi-hop wireless networks \cite{jiang2009distributed}, several congestion control algorithms, \eg, FAST TCP \cite{jin2004fast}, and utility-based resource allocation algorithms in OFDM networks \cite{shams2014survey}). While some of these algorithms have been commercialized, certain others, \eg, TCP Reno and Vegas \cite{brakmo1994tcp}, have been shown to be reverse-engineerable from NUM. Given NUM's success in the classical study of distributed architectures, our aim is to investigate the adaptability of these methods to quantum applications. 
To our knowledge NUM has not been, to date, extended to quantum networks. 

\section{Problem Formulation}
\label{sec:probform}
Classical NUM in its basic form is most applicable under static and continual conditions: \ie, there is a fixed number of user communication sessions, and traffic is injected into the network \textit{ad infinitum}. We will make an analogous assumption for this initial QNUM formulation: namely, that all pairs of users wishing to communicate are fixed, and that each communication session has an infinite backlog of entanglement requests. In principle, there are a number of ways to accommodate changing network dynamics.
In the case of a centralized controller, for instance, one may update and re-solve the optimization program after some time has elapsed; this action could be taken, \eg, after fixed/pre-determined period, whenever an existing communication session drops or a new one arrives, or according to some other heuristic. 
In the case of a completely distributed setting, our proposed QNUM formulation can be distributized (using, \eg, the primal approach after defining suitable cost functions on the constraints of the problem \cite{srikant2013communication}) so that each communication session and each link are responsible for tuning their rate and Werner parameter, respectively, according to feedback received from the network \cite{srikant2013communication}. Another, albeit often less tractable technique, is to transform the problem into a stochastic NUM formulation, see \cite{yi2008stochastic} for an overview. 

As with basic classical NUM, we assume that routes are pre-determined and fixed for all communication sessions, as doing otherwise would require incorporating entanglement routing -- a complex problem in its own right -- into the QNUM formulation. 
A route corresponding to a communication session between two users  in a network is defined by the path connecting the users.
Throughout this manuscript, we often use the terms ``communication session`` and ``route'' interchangeably (similar to the interchangeable use of ``source'' and ``route'' in classical NUM literature). 
The set $\mathcal{R}$ contains all active communication sessions.
The set $\mathcal{L}$ represents the links of a quantum network. We will often use the notation $l\in r$ to denote that link $l$ is a constituent of route $r$. An active communication session experiences QoS both in terms of the rate at which end-to-end entanglement is served to it by the network -- $R_r$ for route $r$, as well as the quality (fidelity) of said states.

In this work, we assume that elementary link-level entanglement is generated using the single-photon scheme. In this scheme, a state of the form
\begin{align}
\rho = (1-\alpha)\ket{\Psi^+}\bra{\Psi^+}+\alpha\ket{\uparrow\uparrow}\bra{\uparrow\uparrow}
\label{eq:ronald_state}
\end{align}
is generated with probability $p_{\text{elem}}=2\eta\alpha$, where $\ket{\Psi^+}$ is a Bell state that is orthogonal to the bright state $\ket{\uparrow\uparrow}$, $\eta$ is the transmissivity between one end of the link and its midpoint (where a midpoint heralding station is located, consisting of a beamsplitter and two detectors), and $\alpha$ is the bright-state population -- a tunable parameter that enables the rate-fidelity tradeoff \cite{humphreys2018deterministic,rozpedek2019building}. We use the following formula for the transmissivity of link $l$ that has length $L_l$:
\begin{align}
\eta_l = 10^{-0.1\beta L_l},
\end{align}
where $\beta=0.2$ dB/km is the fiber attenuation coefficient. In numerical evaluation, we will often multiply $p_{\text{elem}}$ by a constant factor $c\in(0,1)$ to account for various system inefficiencies other than the loss in fiber.

To simplify post-swap fidelity calculations, we assume that instead of the state (\ref{eq:ronald_state}), each link generates Werner states, which may be written as
\begin{align}
\rho_w = w\ket{\Psi^+}\bra{\Psi^+}+(1-w)\frac{\mathbb{I}_4}{4},
\end{align}
parametrized above in terms of the Werner parameter $w$ which relates to $\rho_w$'s fidelity as
\begin{align}
F = \frac{3w+1}{4}.
\label{eq:w_to_fid}
\end{align}
We thus assume that Werner states are generated with success probability $p_{\text{elem}}$ and fidelity $1-\alpha$.
While this approximation will result in lower end-to-end fidelities -- depolarizing noise is considered a worst-case scenario -- it nevertheless allows us to obtain lower-bounds on performance. Moreover, due to the simplicity of the post-swap Werner parameter $w^{\text{e2e}}$, given for a route $r$ by
\begin{align}
w^{\text{e2e}}_r = \prod\limits_{l\in r} w_l,
\label{eq:we2e}
\end{align}
where $w_l$ are Werner parameters of elementary links involved in the swap (see appendices of \cite{inesta2022optimal} for a proof), we can easily construct utility functions with favorable mathematical properties. We finally assume that each link $l$ has an associated repetition time $T_l$ that specifies the time between entanglement generation attempts. We may thus express a link's entanglement generation rate as
\begin{align}
\frac{p_{\text{elem}}}{T} = \frac{2c\eta\alpha}{T} = \frac{2c\eta(1-F)}{T} = \frac{3c\eta}{2T}(1-w)\equiv d(1-w),
\end{align}
where we have defined $d\coloneqq 3c\eta/2T$. We emphasize that there are two different notions of rate in our QNUM formulation: the first is that of a link's entanglement generation rate, given by $d_l(1-w_l)$, and the second is the entanglement rate $R_r$ allotted to route $r$ by the network. A link will subdivide its rate across all routes that it services, and the output of the optimization procedure will determine both the link's rate, as well as its apportionment across active communication sessions. We finally note that configuring the bright-state population parameter of a link is equivalent to configuring the fidelity of the resulting states or, in our approximation, the Werner parameter; we find the latter most convenient to use within the optimization framework.

\section{Quantum Network Utility Maximization Framework}
\label{sec:qnum}
In this section we present a general QNUM framework that can accommodate single-photon entanglement generation schemes and as a result, allows one to explore rate-fidelity tradeoffs found in such quantum networks.
In its most basic form, the QNUM formulation is as follows (for mathematical convenience, instead of maximizing aggregate utility, we minimize its negation):
\begin{align}
\min\quad &-\sum\limits_{r\in \mathcal{R}}U_r(R_r,\vect{w})\\
\text{subject to } &\sum\limits_{r:l\in r}R_r - d_l(1-w_l) = 0,\quad \forall l\in\mathcal{L},\label{constr:rate}\\
&0\leq w_l\leq 1,\quad \forall l\in\mathcal{L}. \label{constr:wbox}
\end{align}
Above, $\vect{w}=\{w_1,\dots,w_{|\mathcal{L}|}\}$ is the vector of all elementary-link Werner parameters.

One may easily incorporate minimum fidelity requirements imposed on end-to-end entangled states generated on routes, by including constraints of the form
\begin{align}
\prod\limits_{l\in r}w_l \geq \frac{4F_r^{\star}-1}{3},\quad \forall r\in \mathcal{R}, \label{constr:e2efid}
\end{align}
where $F_r^{\star}$ denotes the threshold fidelity for route $r$. Note that constraint (\ref{constr:rate}) is equivalent to $\sum\limits_{r:l\in r}R_r = (2c_l\eta_l/T_l)(1-F_l)$, \ie, the aggregate rate at which users/applications receive entanglement at link $l$ is equal to link $l$'s entanglement generation rate.  In principle one could replace the equality with a less-than sign, but we opt for the former to avoid entanglement wastage resulting from a link that produces a surplus of entanglement which then goes unused by the routes. We can convert the inequality constraints (\ref{constr:e2efid}) to a set of convex constraints by taking the logarithm of both sides, to obtain
\begin{align}
\log\left(\frac{4F_r^{\star}-1}{3}\right)-\sum\limits_{l\in r}\log(w_l) \leq 0.
\end{align}
If the utilities $U_r$ are concave, then this formulation results in a convex optimization problem, as the equality constraints (\ref{constr:rate}) are affine (\ie, each is a sum of a linear function and a constant).

Following, we introduce three utility functions which incorporate useful measures of entanglement distribution rate and quality. While two of these measures possess operational meanings, we find that the utility functions which we define on them are not concave in general -- this means that there are few theoretical guarantees for finding global optima. The third utility function on the other hand is concave, albeit it is based on an entanglement measure that is not associated with a significant operational interpretation. While in principle there are an infinite number of quantum network utility functions, we have identified these three as ones that encompass a number of important properties and serve well in identifying favorable utility function behaviors.

\subsection{QNUM based on Distillable Entanglement}
For bipartite states, distillable entanglement can be viewed as the maximum rate at which nearly-perfect Bell states can be distilled from many copies of mixed bipartite states, using local operations and classical communication. While for isotropic\footnote{\textit{I.e.}, states of the form $F\ket{\Psi^+}\bra{\Psi^+}+(1-F)/(d^2-1)(\mathbb{I}-\ket{\Psi^+}\bra{\Psi^+})$.} and Werner states distillable entanglement can be computed numerically, we opt to work here with a lower-bound on this entanglement measure, given by
\begin{align}
D(F) \geq \max\left(1+F\log_2F + (1-F)\log_2\frac{1-F}{3},0\right),
\label{eq:df_lb}
\end{align}
for $F\in[1/2,1]$, see \cite{rains2001semidefinite}. 
Note that the quantity 
\begin{align}
D_H(F) \coloneqq 1+F\log_2F + (1-F)\log_2\frac{1-F}{3},
\label{eq:DH}
\end{align} 
also known as the yield of the so-called ``hashing'' protocol \cite{bennett1996mixed},
 is negative for ${F<0.81}$. To incorporate rate into our utility, we multiply the right-hand side of (\ref{eq:DH}) by $R$, 
and  for a given route $r$, define a utility function as follows:
 \begin{align}
 U^{D}_r(R_r,\vect{w}) = \log\left(R_rD_H(F_r^{\text{e2e}})\right),
 \label{eq:U_D}
 \end{align}
 where $F_r^{\text{e2e}}$ is the end-to-end fidelity of entangled states produced for route $r$.
Above, composition by the $\log$ is motivated by numerous examples encountered in classical NUM wherein such an approach enforces a sense of fairness to all simultaneously active communication sessions found in a network. In particular, since $\lim_{x\to 0}\log(x) = -\infty$, we can be sure that each route receives a fraction of the aggregate distillable entanglement from every link that comprises it. 
A consequence of encasing $D_H$ by the $\log$ is that we implicitly enforce a sufficiently high end-to-end fidelity, \ie, that of at least $\approx 0.81$,
thereby circumventing the restriction that the bound (\ref{eq:df_lb}) holds only for $F\in[1/2,1]$, while also
keeping the formulation simple by not having to set restrictions on the end-to-end fidelity for each route.
The utility function (\ref{eq:U_D}) is very similar to the approach taken by \cite{victora2020purification}, where the authors explored maximization of quantum information flow over repeater chains in a network; notable differences with our approach are 
the restriction of the objective's domain to ensure non-zero distillable entanglement,
and the use of the $\log$ to introduce a notion of fairness.

\subsection{QNUM based on Secret Key Fraction}
Another practical measure of entanglement usefulness is the secret key fraction, which describes the amount of secret key that can be generated from an execution of a QKD protocol (\ie, from a successfully-generated entangled state). We will use the secret key fraction of the BB84 protocol \cite{bennet1984quantum}, which has a convenient closed form for Werner states:
\begin{align}
S_{BB84}(w) = \max\left(1-2h\left(\frac{1-w}{2}\right),0\right),
\label{eq:S_BB84}
\end{align}
where $h(\cdot)$ is the binary entropy function \cite{shor2000simple,lo2005efficient,li2020efficient}. We define a utility function based on (\ref{eq:S_BB84}) as follows:
\begin{align}
U^S_r(R_r,\vect{w}) = \log\left(R_r \left(1-2h\left(\frac{1-w^{\text{e2e}}_r}{2}\right)\right)\right),
\label{eq:U_S}
\end{align}
where $w^{\text{e2e}}_r$ denotes the Werner parameter of an end-to-end entangled state on route $r$.

\subsection{QNUM based on Negativity}
The final entanglement measure which we include in our study is that of negativity \cite{vidal2002computable}, which for a bipartite mixed state $\rho$ on system $AB$ is defined as
\begin{align}
\mathcal{N}(\rho)\equiv \frac{||\rho^{T_A}||_1-1}{2},
\label{eq:negativity}
\end{align}
where $\rho^{T_A}$ is the partial transpose of $\rho$ with respect to subsystem $A$ and $||\cdot||_1$ is the trace norm. $\mathcal{N}(\rho)$ is also equal to the absolute value of the sum of the negative eigenvalues of $\rho^{T_A}$, which reduces to zero for unentangled states.
For a Werner state with fidelity $F$, it can be shown that the negativity corresponds to $F-1/2$, for $F\in[1/2,1]$. As mentioned previously, this quantity does not map cleanly to an operational interpretation; nevertheless, it is one of the easiest entanglement measures to compute. We next define a utility function based on (\ref{eq:negativity}) as follows:
\begin{align}
\tilde{U}^N_r(R_r,\vect{w}) = \log(R_r(F^{\text{e2e}}_r-1/2)),
\end{align}
or equivalently,
\begin{align}
U^N_r(R_r,\vect{w}) = \log\left(R_r\left(3\prod\limits_{l\in r}w_l-1\right)\right),
\label{eq:ngtv_util}
\end{align}
which follows from (\ref{eq:w_to_fid}), (\ref{eq:we2e}), and the fact that we may drop the factor of $1/4$ inside the $\log$ as it has no effect on the optima.

In Appendix \ref{app:negUtilConc}, we show that this utility function is concave -- a feature of great mathematical convenience, as we may employ convex optimization techniques with this formulation to reliably find optima. In contrast, the distillable entanglement- and secret key-based utility functions are not in general concave -- see Appendix \ref{app:distskfNonConc} for proofs. 
\\\\
In all three utility functions, encasing both the rates as well as the entanglement measures within the logarithm is a deliberate design choice, not only for ensuring fairness among the routes, but also for implicitly constraining the domains of these functions so as to ensure non-zero distillable entanglement, secret key rate, and negativity. A consequence, however, is that utility functions (\ref{eq:U_D}) and (\ref{eq:U_S}) have stringent requirements on end-to-end fidelity. As mentioned previously, $D_H$ is negative for fidelities below $0.81$; this implies that any solution to the utility maximization would result in \textit{every} route receiving entanglement of at least this quality, quite likely resulting in low rates $R_r$ across the network. A similar property applies to $U^S$: to see this, let us examine the function $1-2h(p)$, where the relevant interval on $p$ is $[0,1/2]$ since $w\in[0,1]$. This quantity is negative for $p$ less than approximately $0.11$, which translates to a fidelity of $0.84$ -- an even stricter requirement than we had with distillable entanglement. In contrast, the negativity-based utility function merely requires end-to-end fidelities of at least $1/2$. What one would expect, then, is for $U^D$ and $U^S$ to produce higher-quality entanglement at low rates, while the opposite holds for $U^N$. These properties are useful to keep in mind when choosing a utility: the former two, for instance, may be more suitable for QKD or applications with high end-to-end fidelity requirements, while the latter may be more suitable to applications that are sensitive to entanglement inter-arrival times (jitter), such as BQC.


\section{Numerical Investigation}
\label{sec:numexamples}
\begin{figure}[t]
\centering
\includegraphics[width=0.6\linewidth]{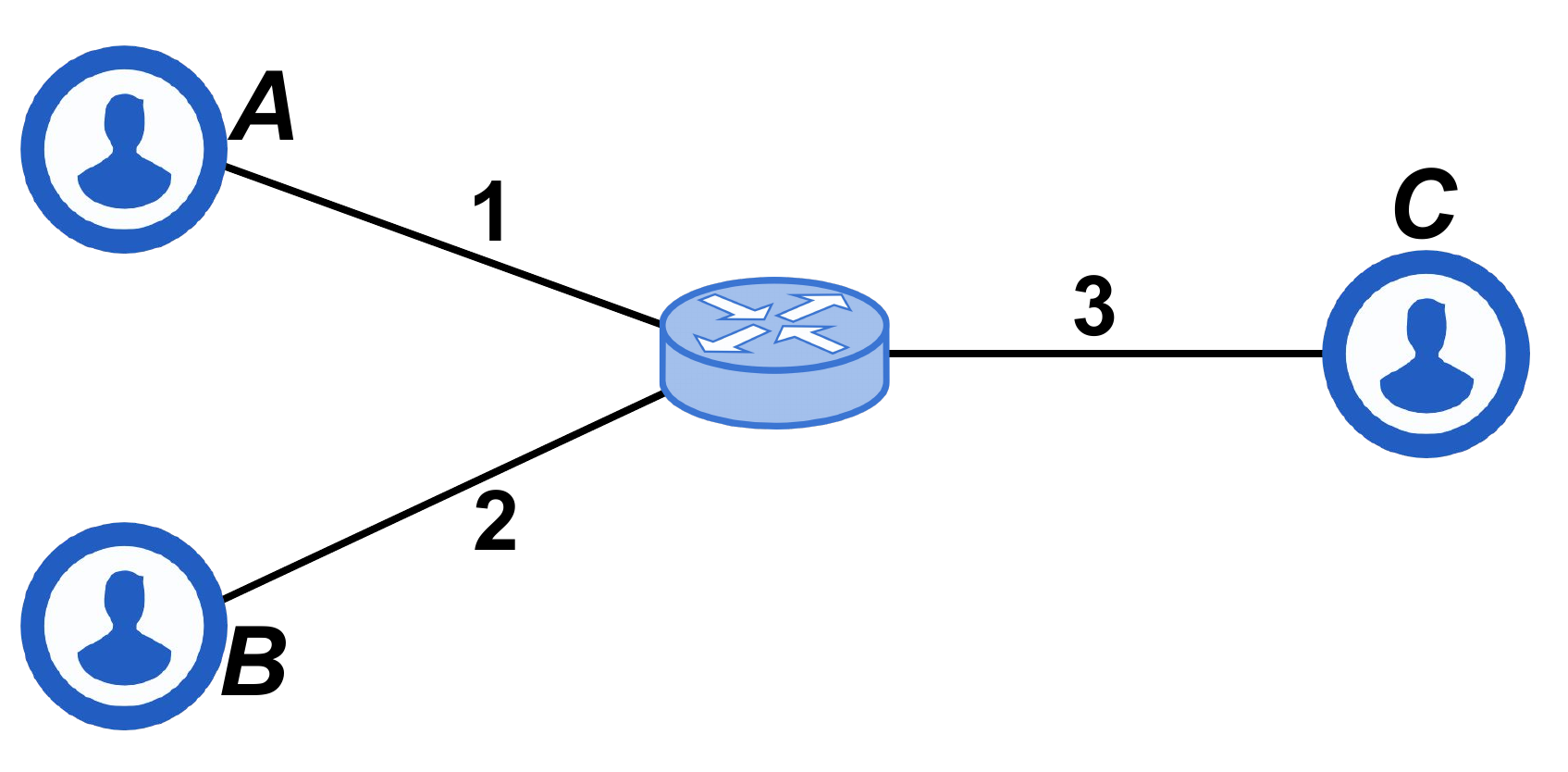}
\caption{A simple four-node, three-link topology; node-pairs $(A,C)$ and $(B,C)$ wish to communicate. For all links, $T=10^{-3}$ s. Links 1 and 2 are 2 km long, while link 3's length is varied in our numerical studies.}
\label{fig:simple3LinkNet}
\end{figure}
We apply the distillable entanglement- (DE), secret key fraction- (SKF), and negativity- (NGTV) based QNUM constructions of the previous section to a number of network topologies chosen so as to expose important behavioral properties of our utility formulations. In particular, we opt for topologies wherein symmetry may be exploited in the sense that all routes derive the same amount of utility from the network -- this allows us to more easily predict and interpret performance metrics of interest. In such highly symmetric network topologies, the optimization is fairly easy to execute via grid search on few parameters (in our examples, only two -- \eg, the Werner parameter of a metropolitan-area link and a backbone link). In general, however, optimization may be performed using more sophisticated, algorithmic methods. For the topology in Figure \ref{fig:simple3LinkNet}, for instance, we use a bound-constrained augmented Lagrangian approach \cite{nocedal2006numerical}
with quadratic penalty functions to enforce the equality constraints (\ref{constr:rate}). For the subproblems, we use the projected gradient descent method with backtracking line search to adaptively choose the step lengths \cite{boyd2004convex}.
\begin{figure*}[t]
\centering
\subfloat{\includegraphics[width=0.32\textwidth,trim={0.5cm 7cm 0.5cm 8cm},clip]{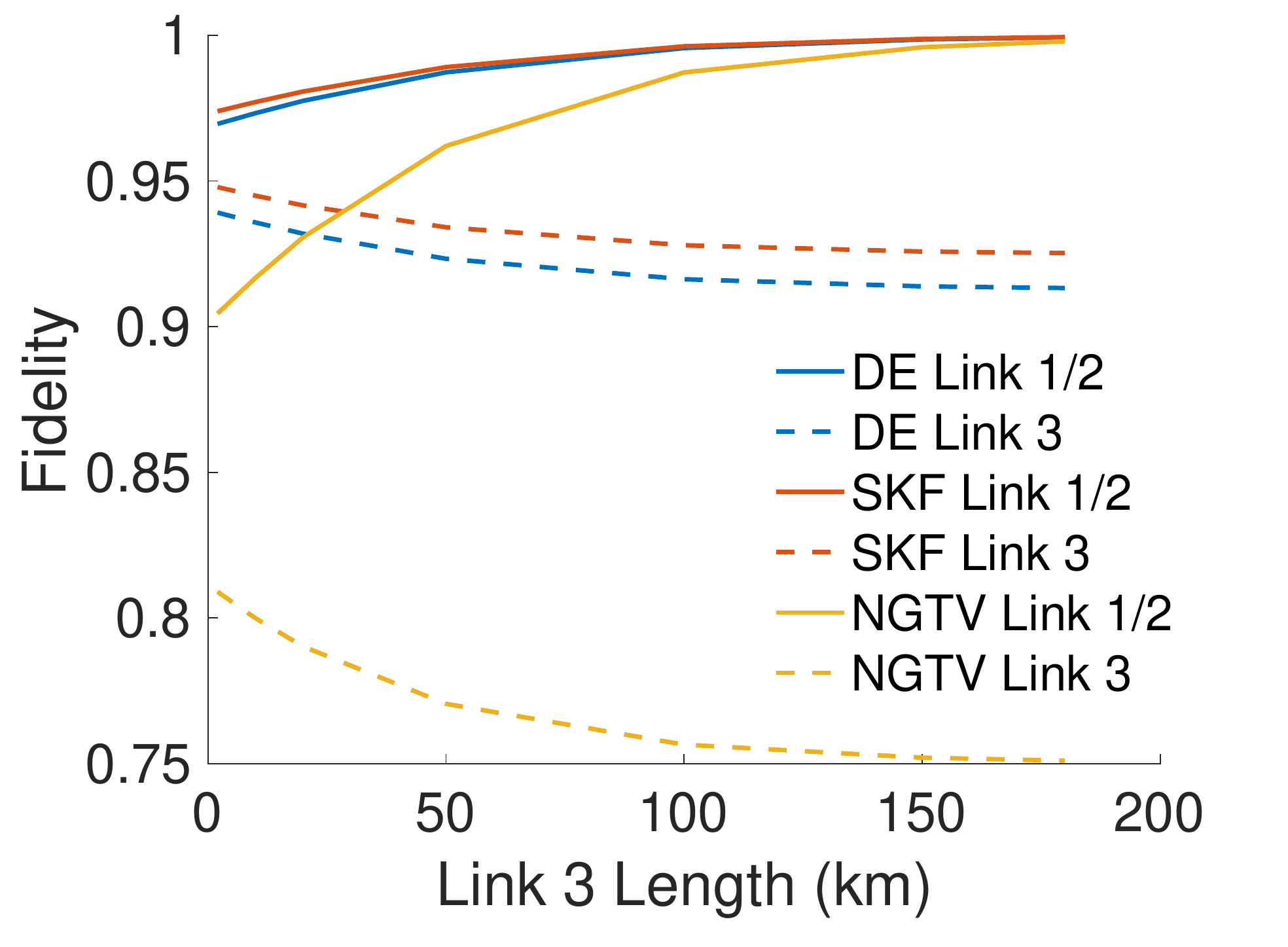}}
\subfloat{\includegraphics[width=0.32\textwidth,trim={0.5cm 7cm 0.5cm 8cm},clip]{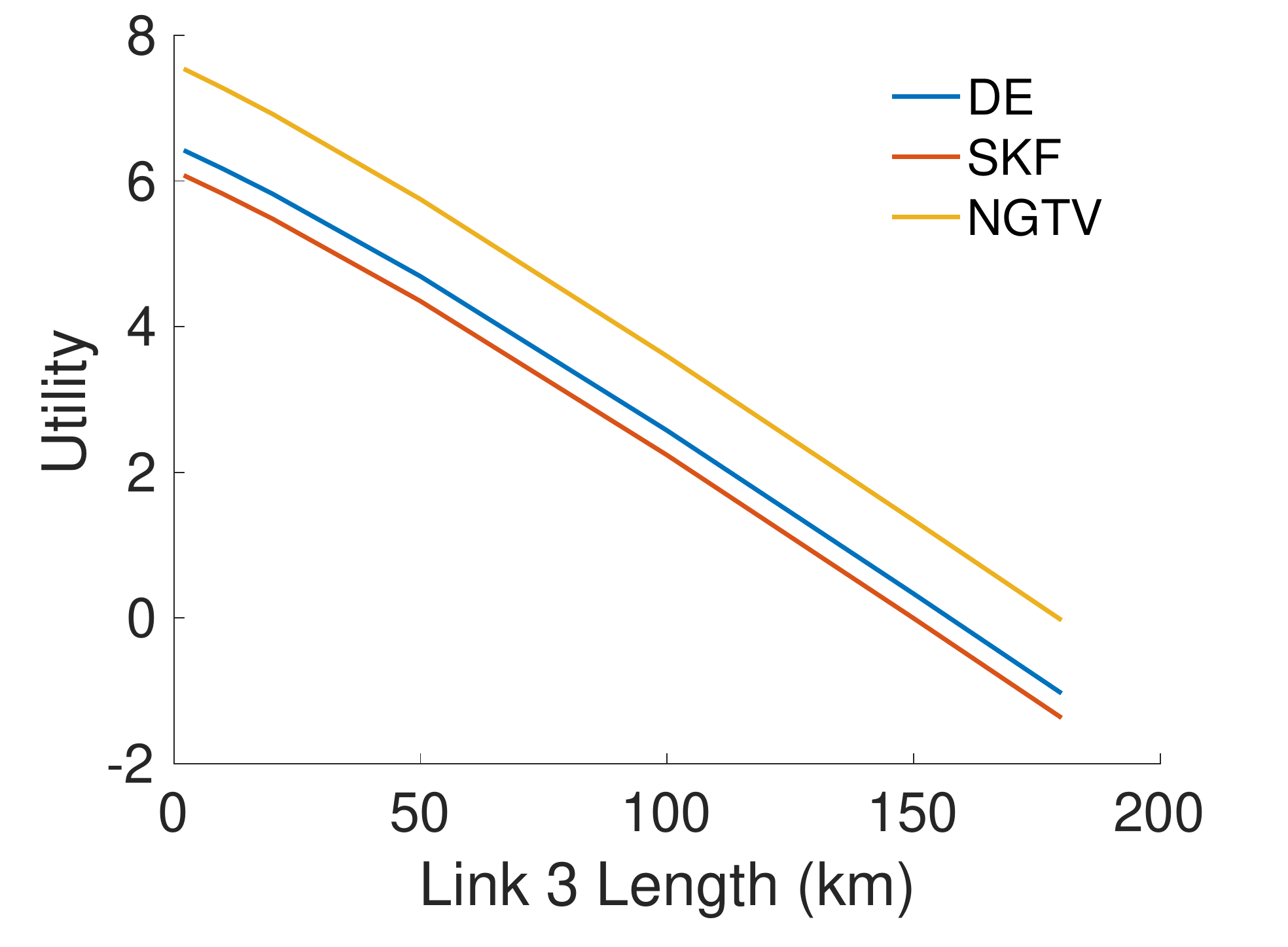}}
\subfloat{\raisebox{-1.3mm}{\includegraphics[width=0.37\textwidth,trim={3cm 2cm 3cm 3cm},clip]{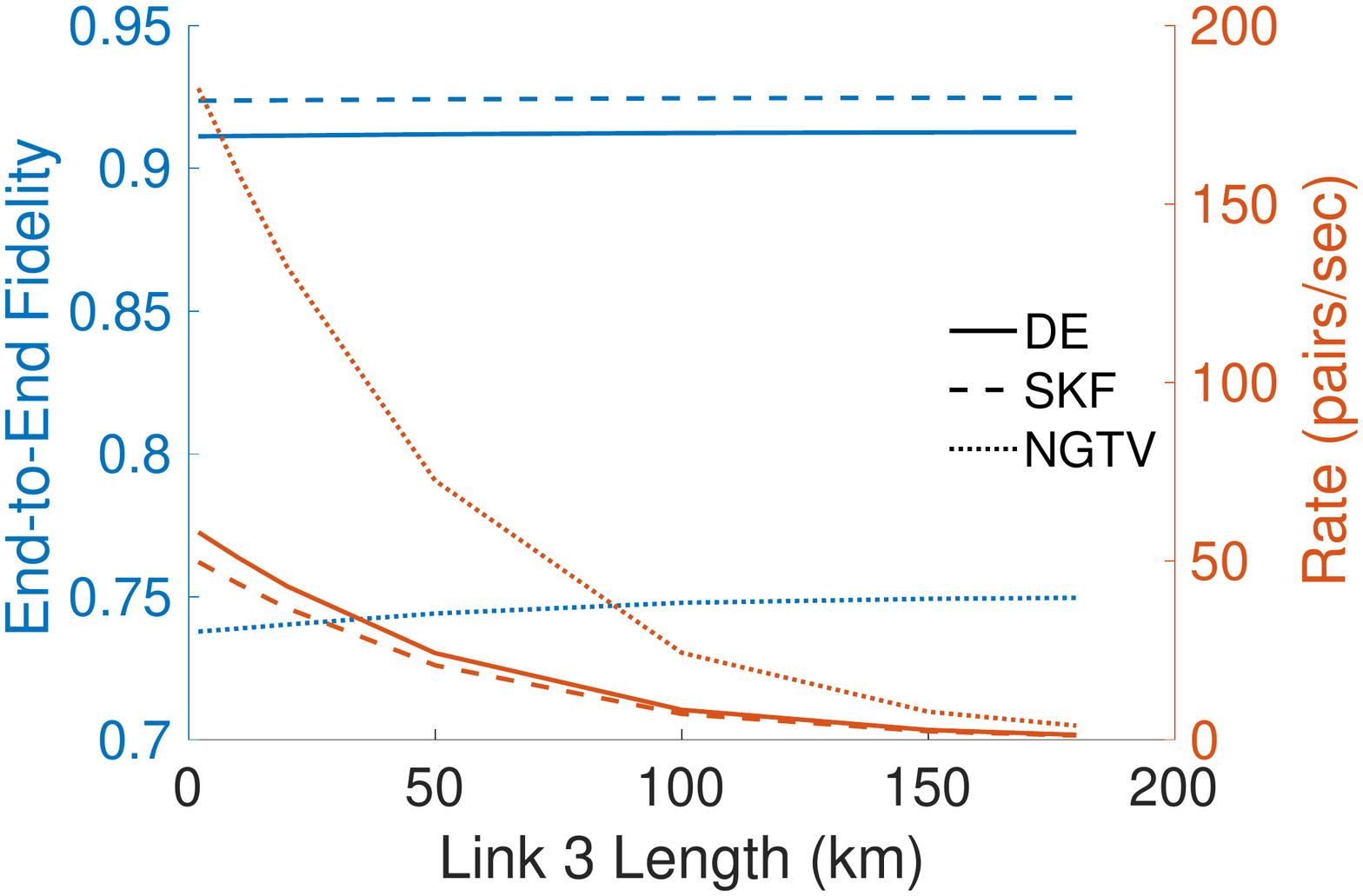}}}
\caption{Optimal link-level fidelity values, resulting utilities, end-to-end fidelities, and rates for the three-link network in Figure \ref{fig:simple3LinkNet}, as a function of the longest link's (link 3) length.}
\label{fig:simple3LinkNetResults}
\end{figure*}

The first topology that we study is depicted in Figure \ref{fig:simple3LinkNet} and is a three-link, three-user network with a single repeater node set up so that the two users on the left wish to communicate with the user on the right. Links 1 and 2 (connecting the users on the left to the repeater) are identical and 2 km long, while link 3's length is varied from 2 to 180 km. All links have constant repetition times of $T=10^{-3}$ s and the non-fiber inefficiencies coefficient $c$ is set to one. There are thus two ``routes'' (communication sessions) in the network, resulting in the following optimization problem:
\begin{align}
&\max~U_1(R_1,\vect{w})+U_2(R_2,\vect{w}) \text{ subject to } \nonumber\\
&R_i = d_{i}(1-w_{i}), i=1,2,\label{eq:3linkNetUtility}\\
&R_1+R_2 = d_3(1-w_3),\nonumber
\end{align}
where the utilities make use of post-swap Werner parameters of route 1/2, which are given by $w_1w_3/w_2w_3$, respectively. Figure \ref{fig:simple3LinkNetResults} (left) presents individual elementary link-level fidelities for this topology. As one would expect, since link 3 must service two communication sessions, while links 1 and 2 service only one each, the former's fidelity is in general lower in order to support the additional rate demand. Links 1 and 2, on the other hand, have higher fidelities to compensate for link 3's lower one. Moreover, as link 3's length increases, links 1 and 2 are further coerced to produce higher-quality entanglement to make up for the loss in fidelity of link 3's entanglement.

The resulting aggregate network utility (Figure \ref{fig:simple3LinkNetResults} center) decreases approximately linearly as a function of link 3's length, evidently due to the network's lowered capability of producing entanglement of high-quality, or at high rates. The rightmost panel of the figure presents end-to-end fidelities and rates as a function of link 3's length, for each individual route (as previously mentioned, all routes receive the same QoS due to the symmetry of the problem).
We observe here that all three utilities appear to maintain a fairly steady end-to-end entanglement fidelity, while sacrificing route rates, which appear to decrease at an exponential rate with the link length.

A striking observation so far is that DE- and SKF-based utility formulations appear to be in close agreement -- they produce similar solutions and subsequently have comparable notions of utility. The negativity-based formulation, on the other hand, puts rate and fidelity on a much more equal footing, thereby allowing considerably lower end-to-end fidelities. Of the three utilities, the SKF-based one is most stringent on fidelity (albeit only by a small amount relative to the DE-based utility).

\begin{figure}[t]
\centering
\includegraphics[width=0.98\linewidth]{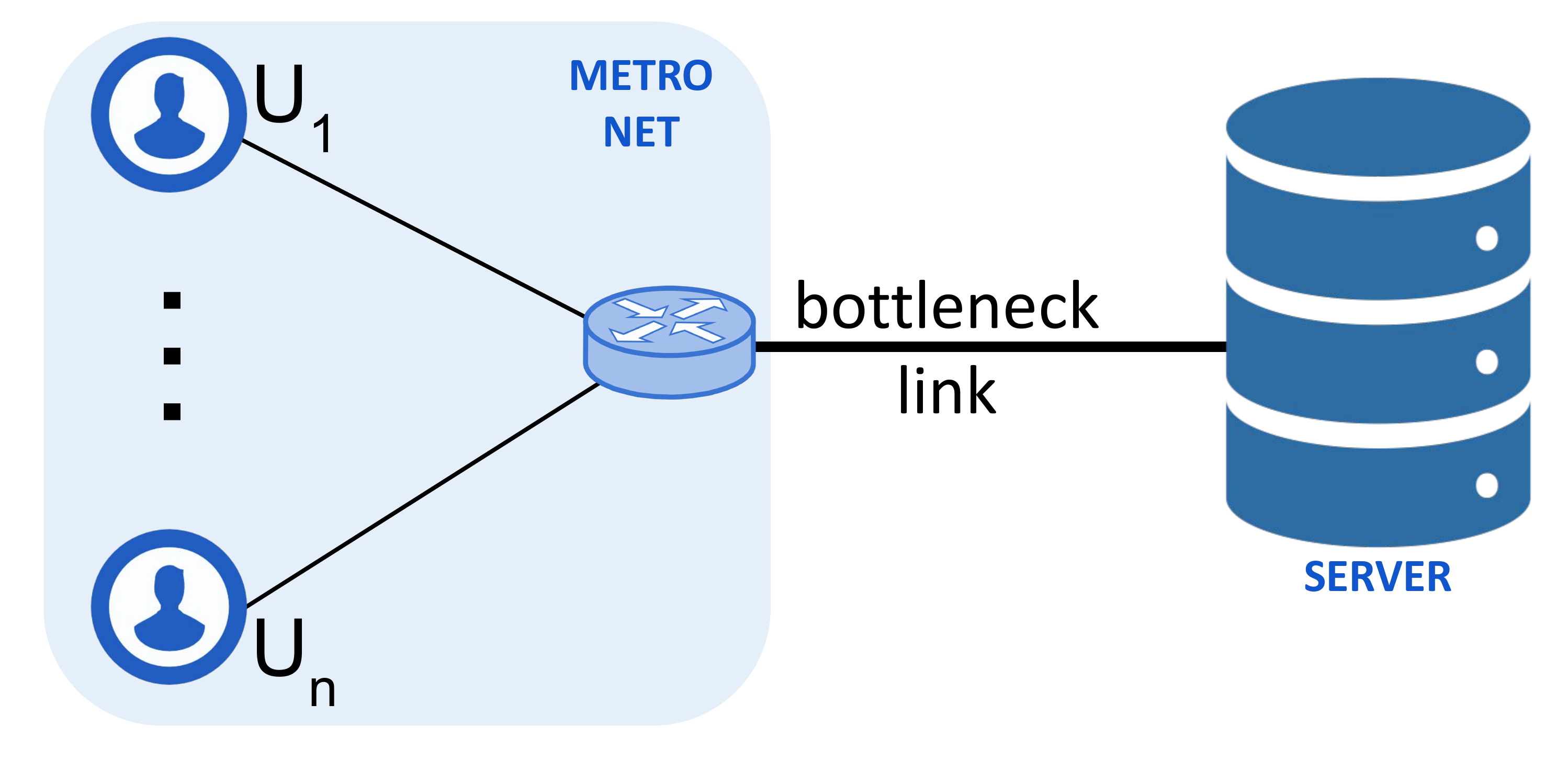}
\caption{A clients-server topology with $n$ users and one repeater node connecting them to the server via a bottleneck link. Links that connect users to the network are identical and 15 km with $T=10^{-3}$ s, while the bottleneck link is 100 km with $T=10^{-4}$ s. The number of users is varied in our numerical studies.}
\label{fig:clientServer}
\end{figure}
\begin{figure*}[t]
\centering
\subfloat{\includegraphics[width=0.32\textwidth,trim={0.5cm 7cm 0.5cm 8cm},clip]{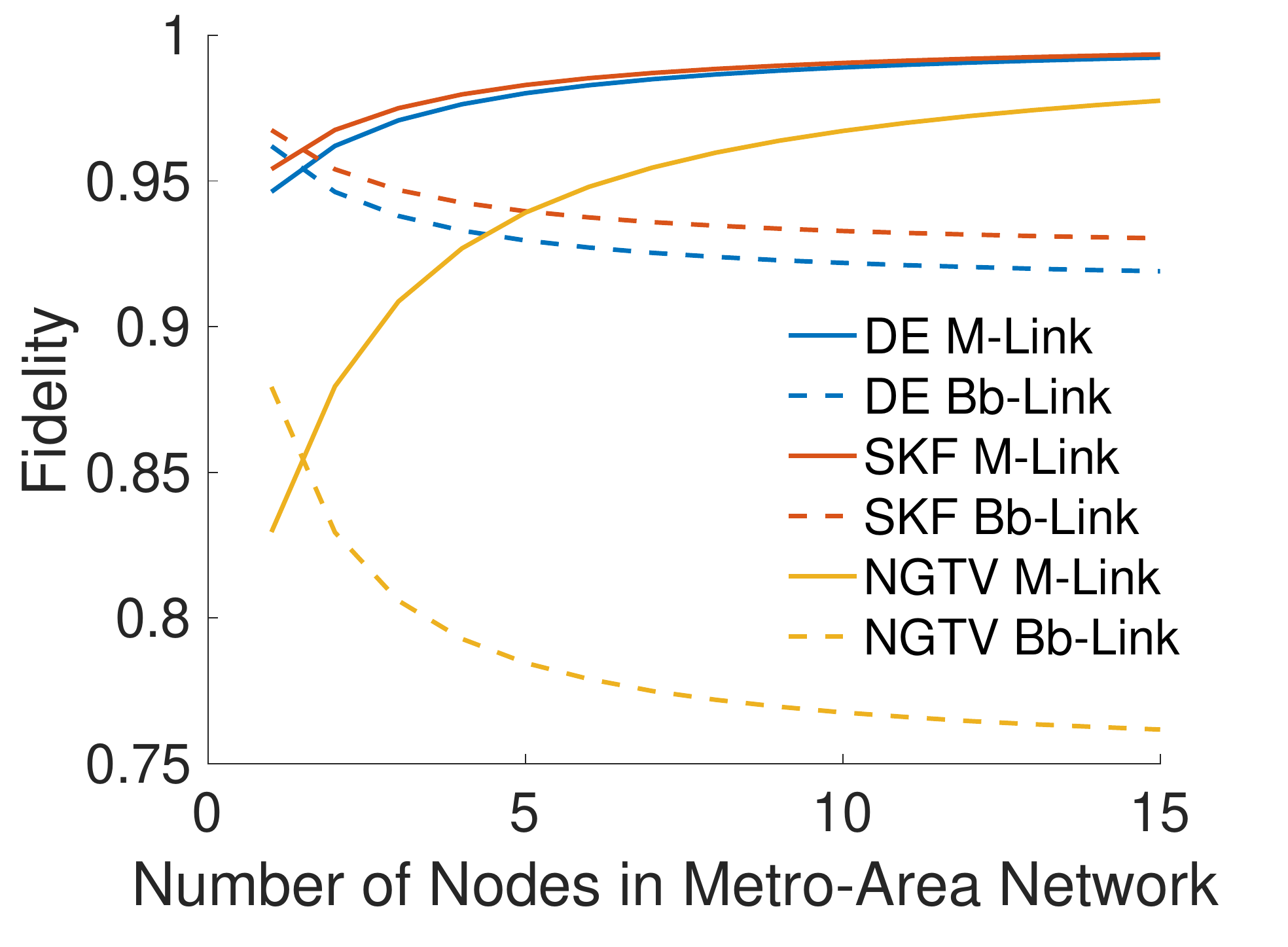}}
\subfloat{\includegraphics[width=0.32\textwidth,trim={0.5cm 7cm 0.5cm 8cm},clip]{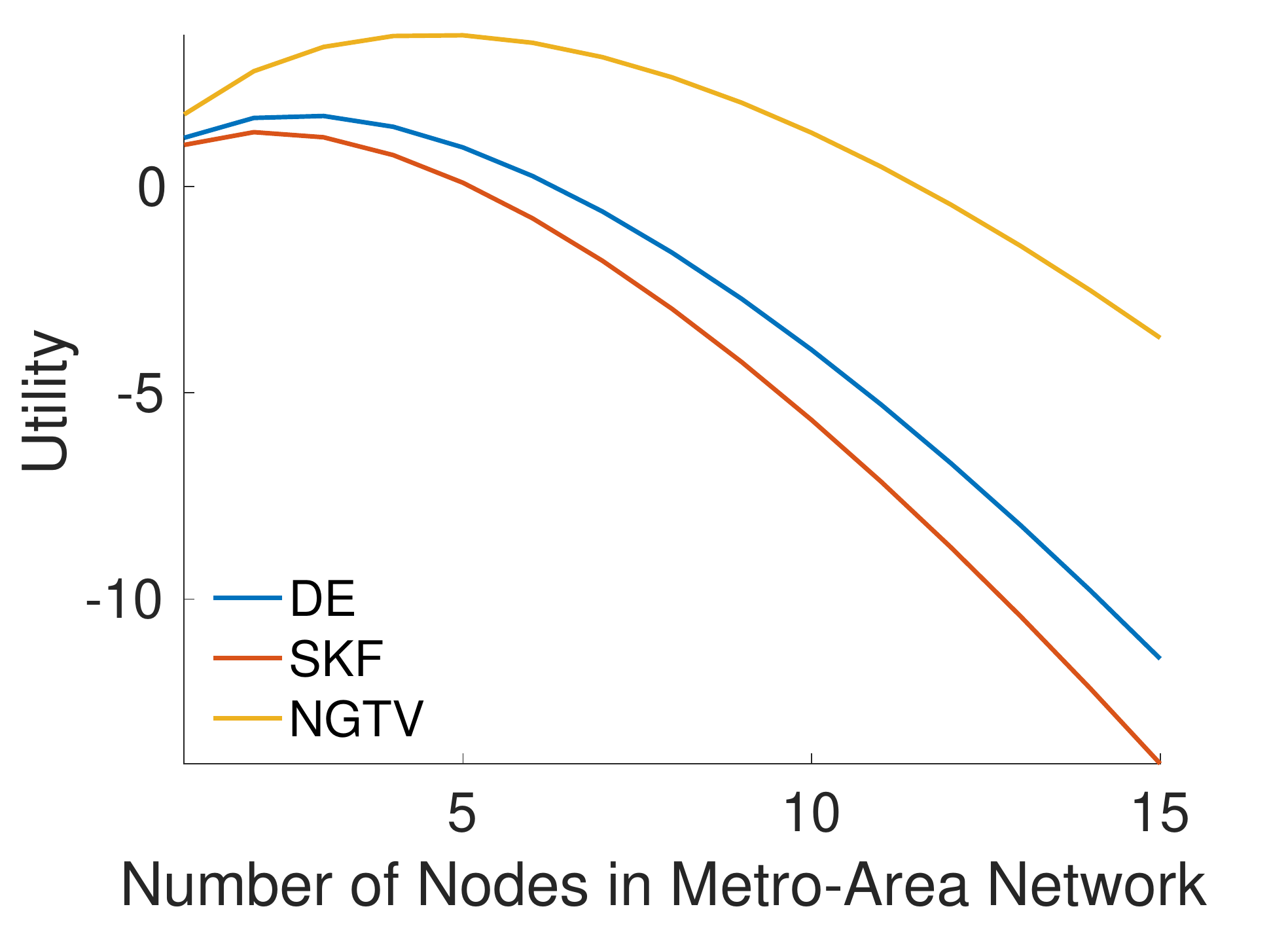}}
\subfloat{\raisebox{-1.3mm}{\includegraphics[width=0.37\textwidth,trim={3cm 2cm 3cm 3cm},clip]{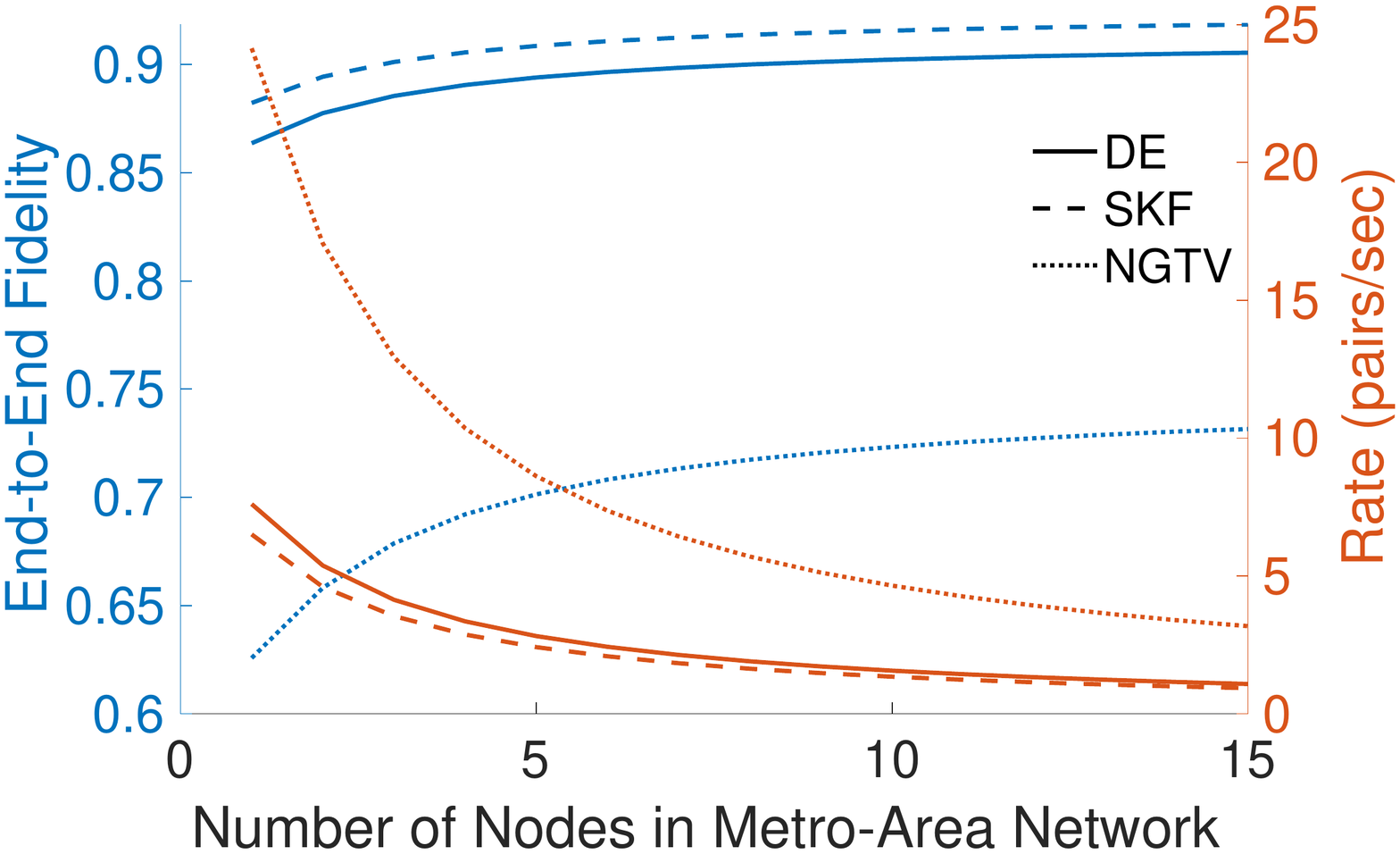}}}
\caption{Optimal link-level fidelity values, resulting utilities, end-to-end fidelities, and rates for the client-server network in Figure \ref{fig:clientServer}, as a function of the number of users.}
\label{fig:clientServerResults}
\end{figure*}

A generalization of the topology in Figure \ref{fig:simple3LinkNet} is the ``clients-server'' topology illustrated in Figure~\ref{fig:clientServer}. Here, multiple users connected to a repeater node form a rudimentary metropolitan-area network. All user-repeater links are identical, have 15 km lengths, and repetition times $T=10^{-3}$ s. The repeater-server link, which has the potential to become a bottleneck, is a longer-distance link of 100 km and has $T=10^{-4}$ s. For all links, $c=0.1$. The QNUM formulation is similar to that of (\ref{eq:3linkNetUtility}), with the exception that we sum over all $n$ route utilities within the objective, and the repeater-server link (denoted by the subscript $\text{Bb}$) yields a constraint of the form
\begin{align*}
\sum\limits_{i=1}^n R_i = d_{\text{Bb}}(1-w_{\text{Bb}}).
\end{align*}

Figure~\ref{fig:clientServerResults} (left) presents the optimal link-level fidelities as a function of the number of users in the metro-area network, with ``M-Link'' denoting metropolitan-area links and ``Bb-Link'' denoting the repeater-server link. When there is only one user in the metro-area network, the repeater-server link is relatively unburdened, and due to $d_{\text{Bb}}>d_{\text{metro}}$, where $d_{\text{metro}}$ is any metropolitan-area link's parameter, produces higher-fidelity entanglement than metro-area links. As the number of nodes grows, increasing the load on ``Bb-link'', we observe a crossover -- the link's fidelity must decrease to accommodate the additional demand, causing metro-area links to compensate in fidelity.

Figure~\ref{fig:clientServerResults} (center) presents the resulting network utility, where we observe a new behavior: up to a threshold, the aggregate utility increases with the number of nodes. A possible explanation for this is that for a sufficiently small number of nodes, the network may be underutilized. Consider a scenario where a powerful backbone link, for example, is able to generate high-fidelity entanglement at a relatively high rate. An individual route that utilizes such a link may not be able to keep up with this rate of entanglement generation without excessively impairing the fidelity of the states it generates. If the number of routes utilizing the backbone link increases (up to a threshold which depends on the utility function and other network parameters), the aggregate utility may increase even if individual route utilities are lower than the utility of the single-route scenario. Beyond the threshold, the backbone link becomes overburdened and begins to sacrifice fidelity, resulting in lower aggregate utilities.

Figure~\ref{fig:clientServerResults} (right) presents individual route end-to-end fidelities and rates -- as with the topology example in Figure~\ref{fig:simple3LinkNet}, we observe rapid decreases in the rates when resources become scarcer in the network (\ie, as the number of users increases in the present example). As the repeater-server link begins to service more users, thereby splitting its aggregate entanglement generation rate among a larger number of routes, we also observe an increase in end-to-end fidelity -- a consequence of metro-area links being coerced to generate entanglement at lower rates. 

\begin{figure}[t]
\centering
\includegraphics[width=0.98\linewidth]{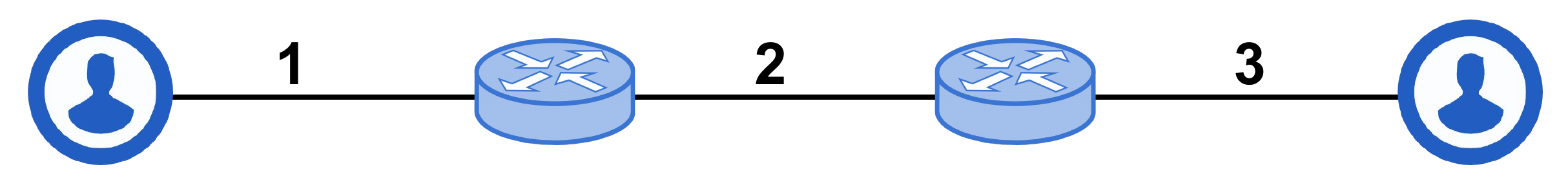}
\caption{A line topology with two users and two repeater nodes. Links 1 and 3 are identical, while link 2's length is varied for our numerical studies.}
\label{fig:LineNet}
\end{figure}
\begin{figure*}
\centering
\subfloat{\includegraphics[width=0.32\textwidth,trim={0.5cm 7cm 0.5cm 8cm},clip]{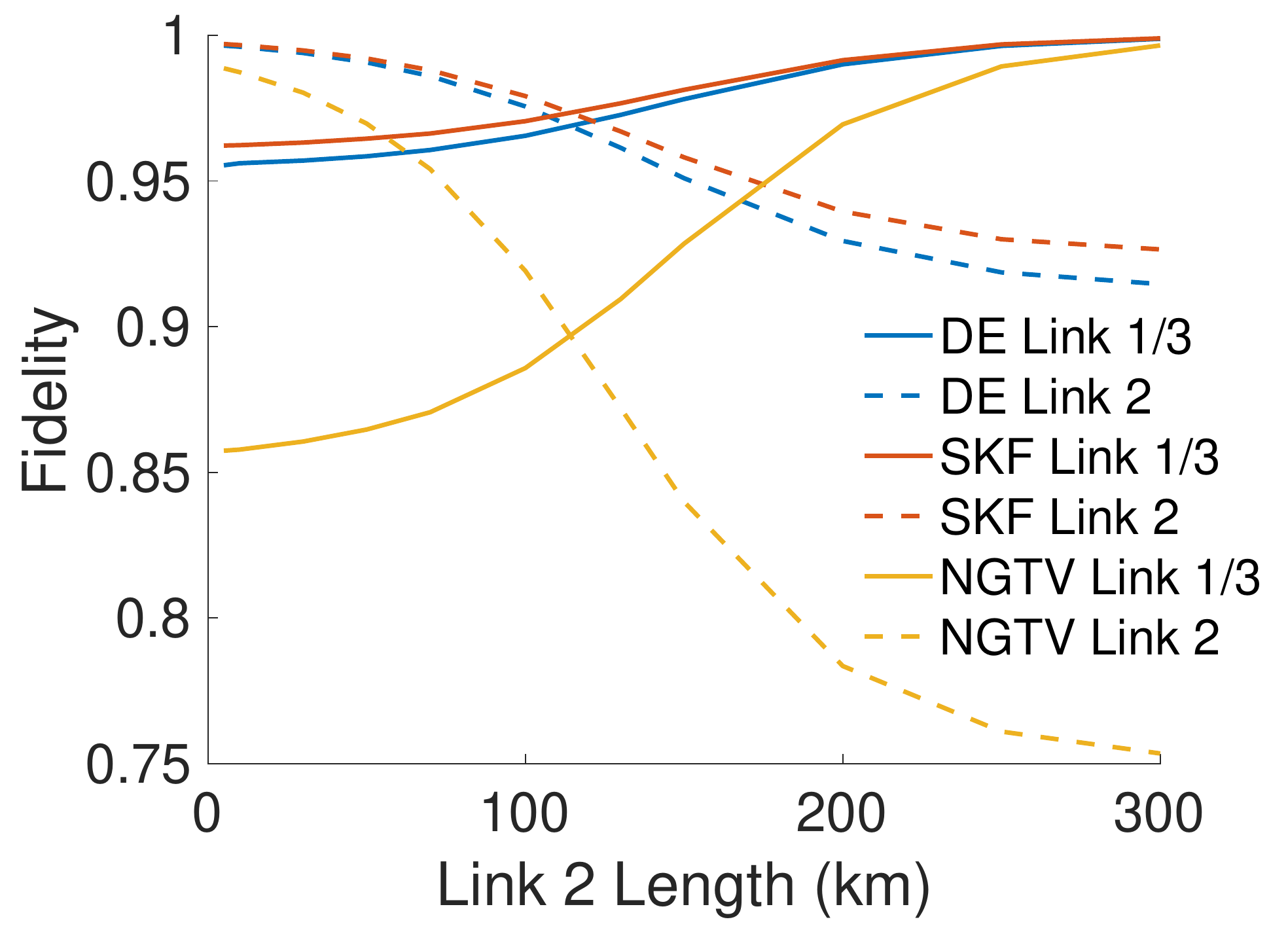}}
\subfloat{\includegraphics[width=0.32\textwidth,trim={0.5cm 7cm 0.5cm 8cm},clip]{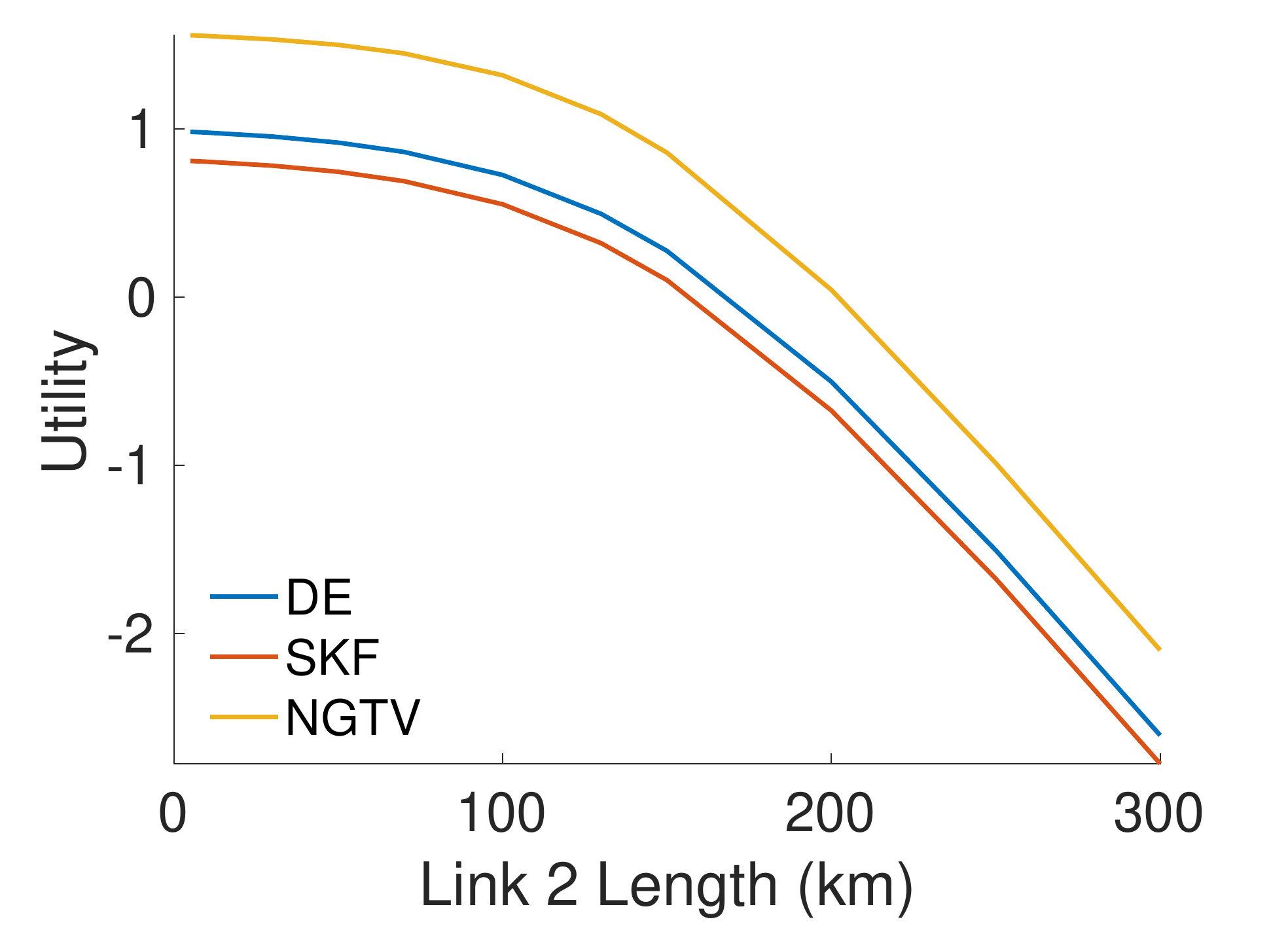}}
\subfloat{\raisebox{0.8mm}{\includegraphics[width=0.37\textwidth,trim={4cm 3cm 3cm 3cm},clip]{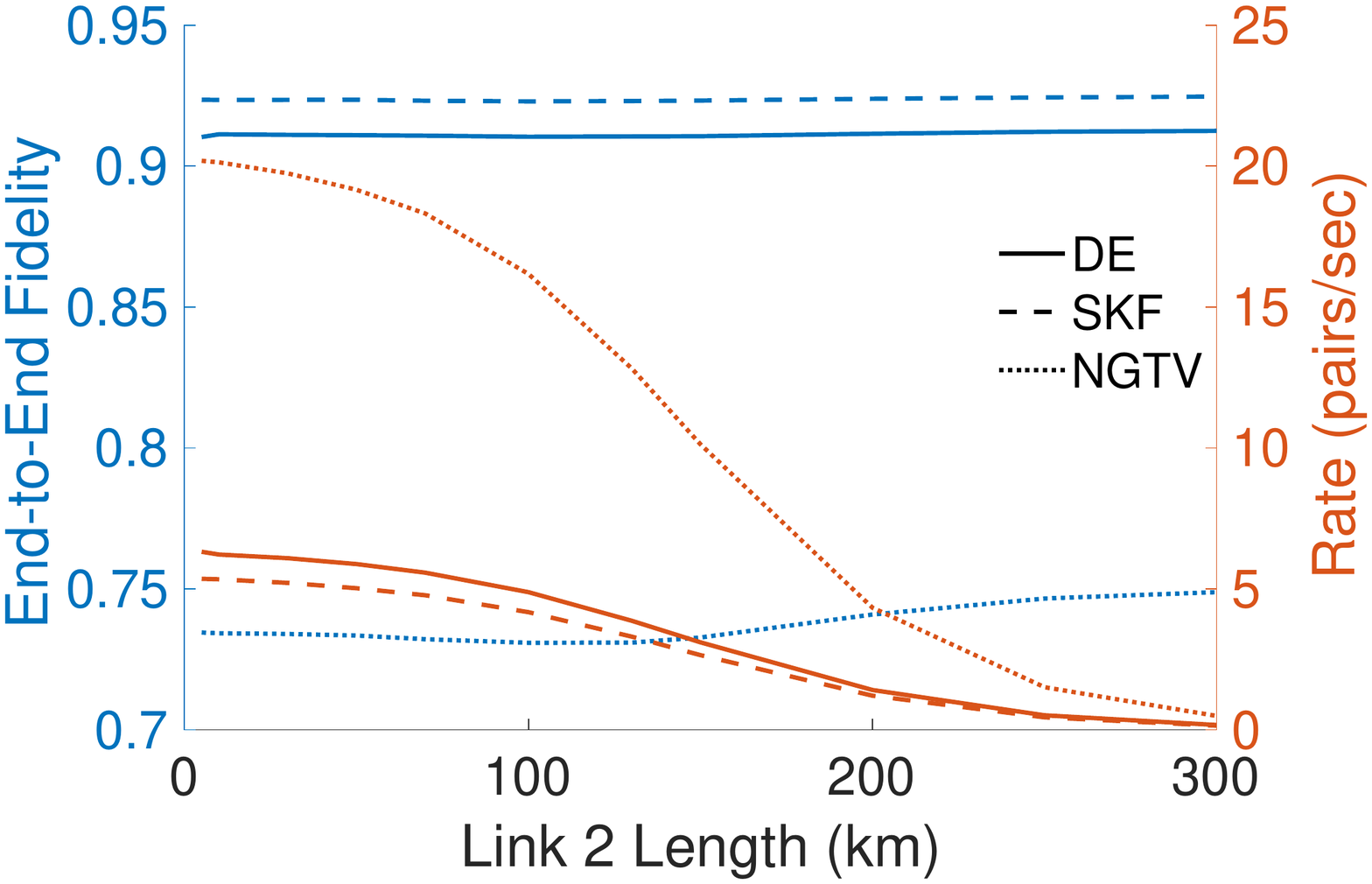}}}
\caption{Optimal link-level fidelity values, resulting utilities, end-to-end fidelities, and rates for the line network in Figure~\ref{fig:LineNet}, as a function of the middle link's (link 2) length.}
\label{fig:LineNetResults}
\end{figure*}

We next study a simple line network is shown in Figure~\ref{fig:LineNet}. Here, links 1 and 3 are both 15 km and have $T=10^{-3}$ s, while link 2 has $T=10^{-4}$ s and its length is varied between 5 and 300 km. For all links, $c=0.1$. Figure~\ref{fig:LineNetResults} (left) presents the optimal link-level fidelity as a function of link 2's length, where the crossover effect is even more apparent than that of the previous example. In the center panel of the figure, we observe a monotonic decrease in the network utility -- a natural consequence of link 2's decreased ability of producing high-fidelity states. In the right panel of the figure, we again see little deviation in an individual route's end-to-end fidelity, while rates now exhibit a more interesting behavior -- both concave as well as convex profiles in the rate curves.

\begin{figure}[t]
\centering
\includegraphics[width=0.98\linewidth]{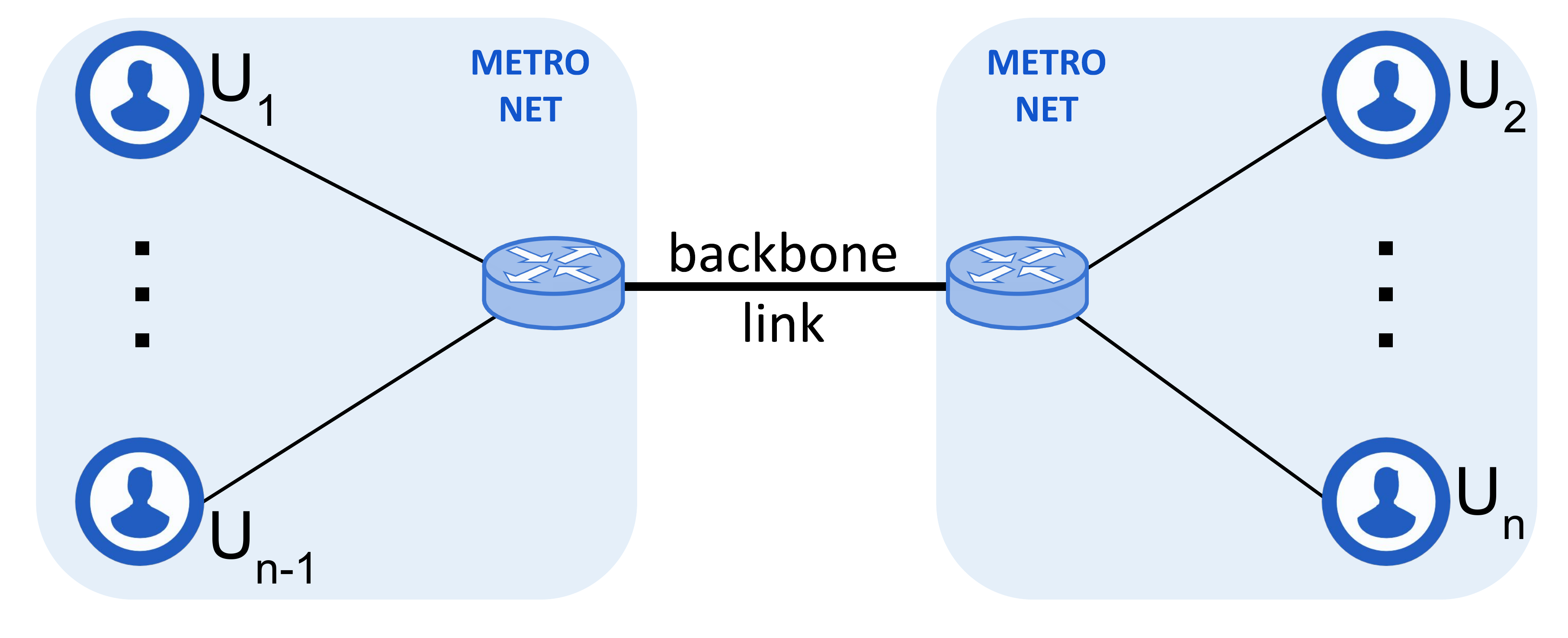}
\caption{A dumbbell topology with $n$ users and two repeater nodes connected by a backbone link. Links that connect users to the network are identical. User pairs $(U_1,U_2),\dots,(U_{n-1},U_n)$ wish to communicate. The number of communicating user pairs is varied in our numerical studies.}
\label{fig:DumbbellNet}
\end{figure}
\begin{figure*}[t]
\centering
\subfloat{\includegraphics[width=0.32\linewidth,trim={0.5cm 7cm 1cm 8cm},clip]{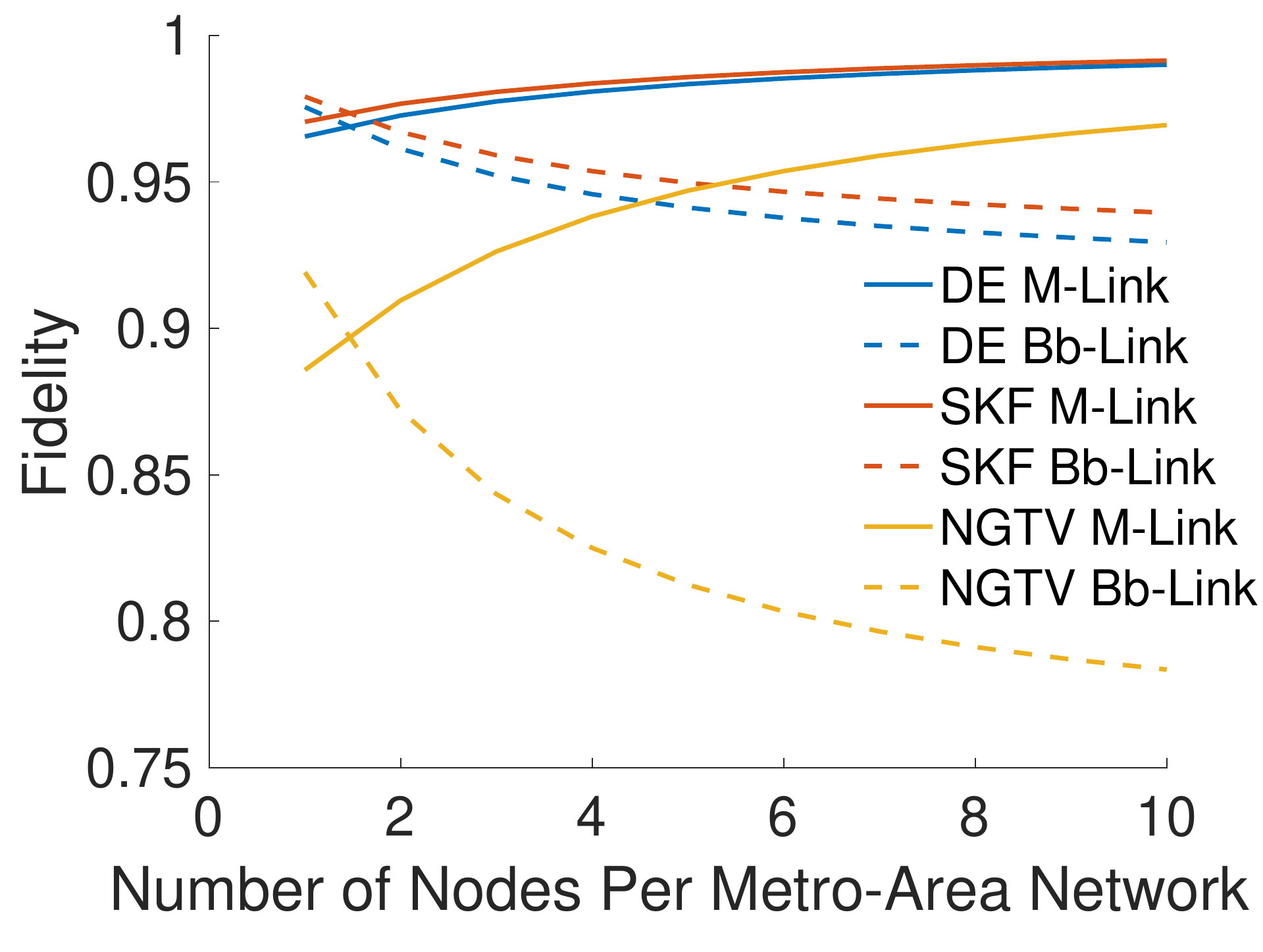}}
\subfloat{\includegraphics[width=0.32\linewidth,trim={0.5cm 7cm 1cm 8cm},clip]{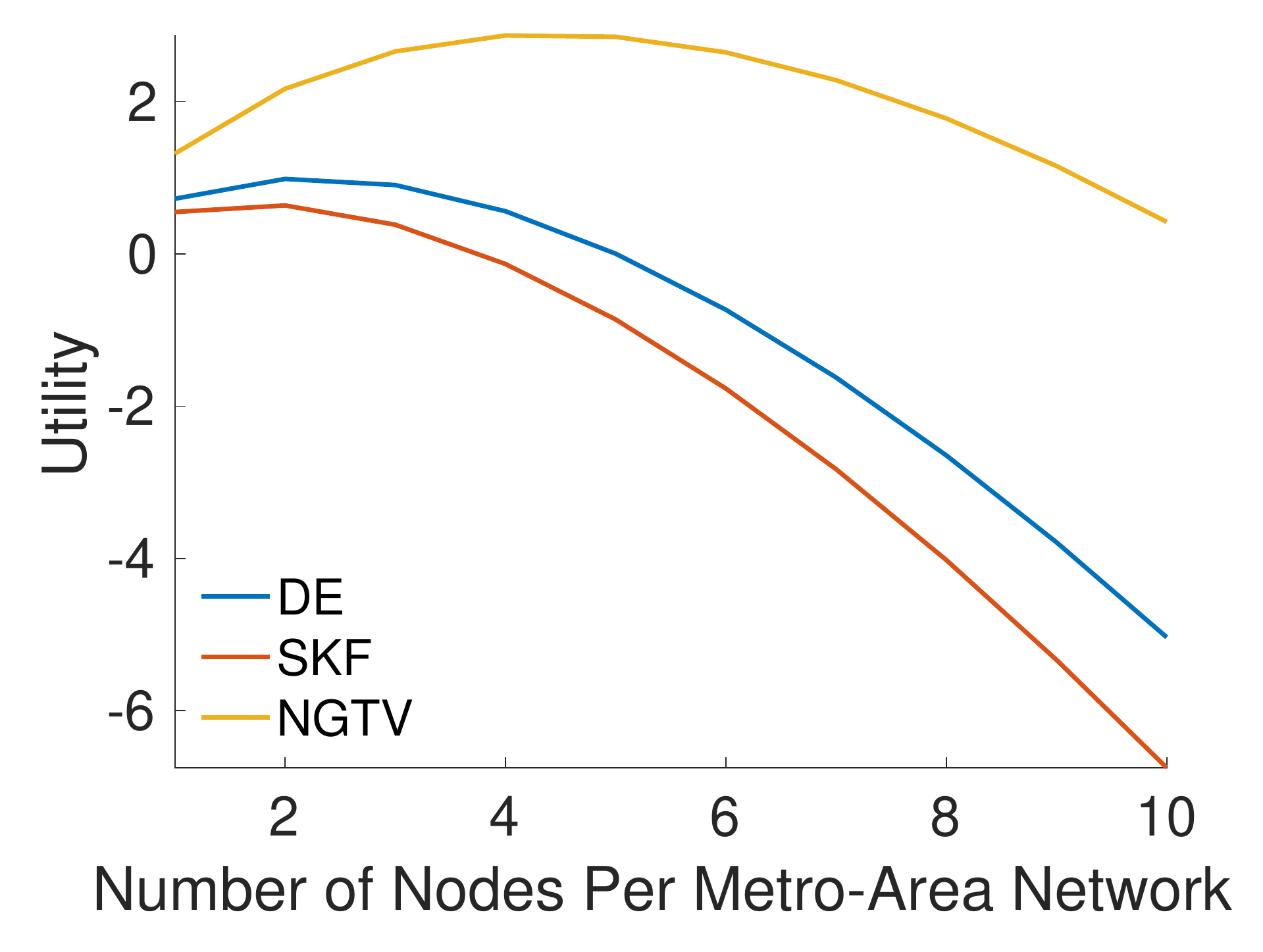}}
\subfloat{\raisebox{1.1mm}{\includegraphics[width=0.37\linewidth,trim={3.5cm 3cm 3.5cm 3.9cm},clip]{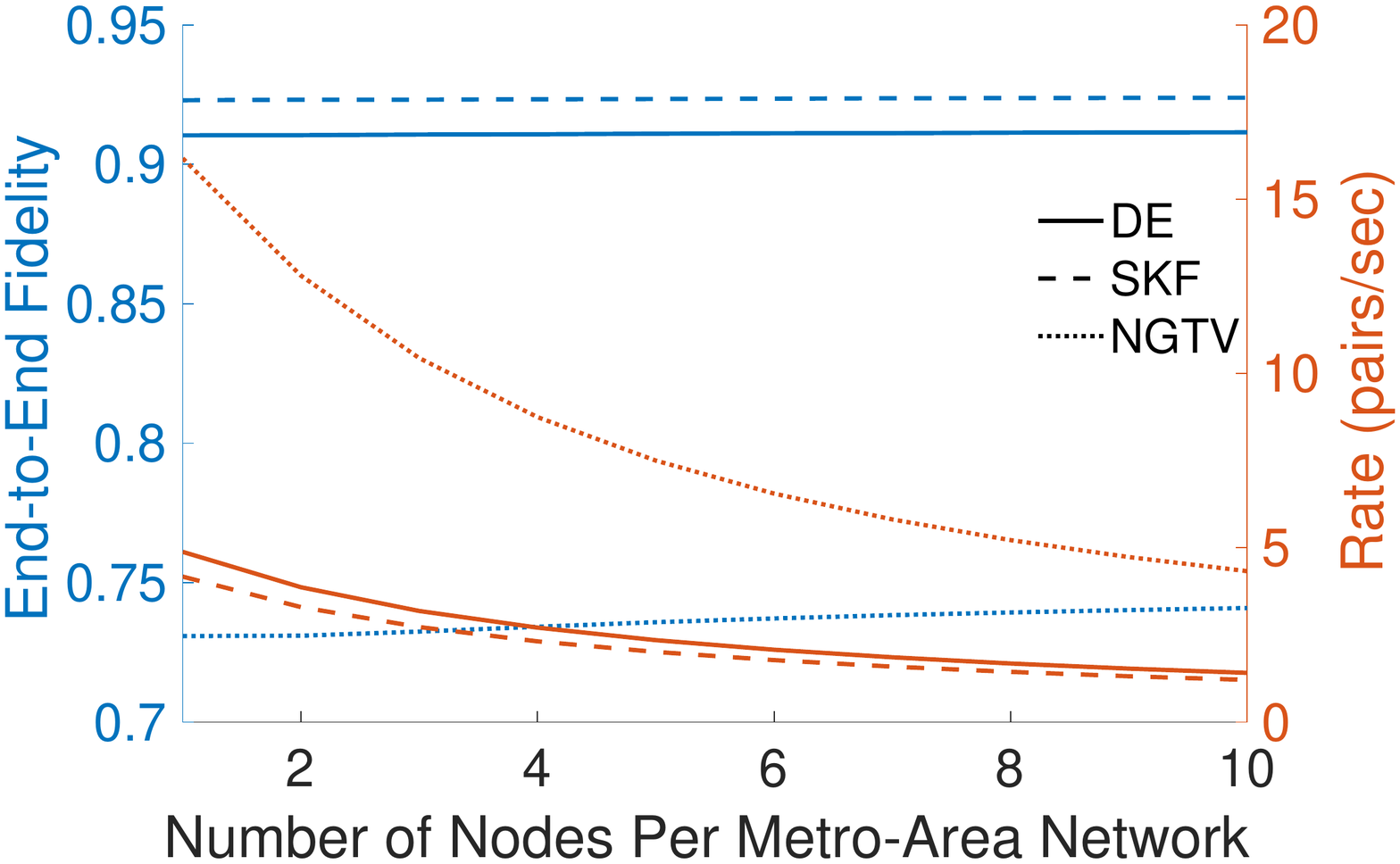}}}
\caption{Optimal link-level fidelity values, resulting utilities, end-to-end fidelities, and rates for the dumbbell network in Figure \ref{fig:DumbbellNet}, as a function of the number of nodes in each metropolitan-area network.}
\label{fig:DumbbellNetResults}
\end{figure*}

Our final numerical study is that of the dumbbell topology as illustrated in Figure~\ref{fig:DumbbellNet}. Here, each metropolitan-area network has $n/2$ users, each of which has a communication session with a user in the other network. A backbone link connects the two metropolitan areas. All metro-area links are identical 15 km links with $T=10^{-3}$ s, while the backbone is a 100 km link with $T=10^{-4}$ s. $c=0.1$ for all links. Figure~\ref{fig:DumbbellNetResults} (left) presents the optimal metro-area (``M-Link'') and backbone (``Bb-Link'') fidelities, as the number of nodes in each metro-area varies. We again see the crossover effect, as well as the increase in aggregate utility (up to a threshold) similar to that of the clients-server network example (center panel). Finally, the right panel of the figure presents a familiar behavior: fairly constant end-to-end fidelities, with rates paying the price.

\section{Conclusion}
\label{sec:concl}
The aim of this work has been to explore ways of adapting classical NUM to quantum networked settings. To this end, we proposed three candidate quantum network utility functions, each based on a different entanglement measure. We applied QNUM to a number of network topologies to solve the resource allocation problem for networks that have inherent rate-fidelity tradeoffs during the elementary link entanglement generation process.
We have found that formulations incorporating distillable entanglement and secret key fraction are quite similar in their behavior, and that they both place a high value on end-to-end fidelity of entanglement. Our third formulation, based on entanglement negativity, allows for lower end-to-end fidelities, and as a result yields higher rates to user-pairs wishing to share entanglement. These contrasting behaviors raise questions about suitable QNUM constructions tailored to specific distributed quantum applications.

\appendices
\section{Concavity of the Negativity-Based Utility Function}
\label{app:negUtilConc}
In this appendix, our goal is to show that the utility function $U_r^N(R_r, \vect{w}) = \log\left(R_r\left(3\prod\limits_{l\in r}w_l-1\right)\right)$ defined on the domain
\begin{align}
\left\{R_r>0, \vect{w} > \vect{0}: \prod\limits_{l\in r}w_l >\frac{1}{3}\right\}
\label{eq:UNdom}
\end{align}
 is concave.
To do so, it suffices to show that the function $f(\vect{x}) = -\log(\prod\limits_i x_i -1)$, where $x_i>0$ are such that $\prod x_i > 1$, is convex. The reason for this is as follows:
\begin{align}
\log\left(R_r\left(3\prod\limits_{l\in r}w_l-1\right)\right) 
&= \log R_r + \log\left(\prod\limits_{l\in r}\sqrt[|r|]{3}w_l-1\right),
\end{align}
where $|r|$ is the number of links in route $r$. It is known that a sum of concave functions is also concave. The argument then follows by letting $x_l \equiv \sqrt[|r|]{3}w_l$, where we know that $x_l> 0$ since $w_l$ are strictly positive, and we also know that $\prod\limits_l x_l >1$ since by (\ref{eq:UNdom}), $3\prod\limits_l w_l >1$.

Our first goal is to prove that the domain of this function, \ie, the set $S\equiv \left\{\vect{x}>\vect{0}:\prod x_i > 1\right\}$, is a convex set. For this we must ensure that given any two points $\vect{x}\neq \vect{y} \in S$, $\theta \vect{x}+(1-\theta)\vect{y}$ is also in $S$, for any $\theta \in [0,1]$. \textit{I.e.}, we require
\begin{align}
\prod\limits_{i=1}^n (\theta x_i+(1-\theta) y_i) >1,
\label{eq:conv_req}
\end{align}
where $n$ is the number of elements in $\vect{x}$ (or $\vect{y}$). Expanding the left-hand side of (\ref{eq:conv_req}), we have
\begin{align}
&\theta^n\prod\limits_{i=1}^n x_i + (1-\theta)^n \prod\limits_{i=1}^n y_i + \nonumber\\
&\sum\limits_{k=1}^{n-1}\theta^k (1-\theta)^{n-k} \sum\limits_{i_1=1}^{n-k+1}\hspace{-1mm}\sum\limits_{i_2=i_1+1}^{n-k}\hspace{-2mm}\cdots\hspace{-2mm}\sum\limits_{i_k=i_{k-1}+1}^{n}\hspace{-3mm}x_{i_1}x_{i_2}x_{i_k}\hspace{-2mm}\prod\limits_{j\neq i_1,\dots,i_k}\hspace{-2mm}y_j\nonumber\\
&>\theta^n\prod\limits_{i=1}^n x_i + (1-\theta)^n \prod\limits_{i=1}^n y_i+ \nonumber\\
&\sum\limits_{k=1}^{n-1}\theta^k (1-\theta)^{n-k} {n \choose k}\left(\prod\limits_{i=1}^n x_i^{{n-1\choose k-1}}\prod\limits_{j=1}^n y_j^{n-1\choose k}\right)^{1/{n \choose k}},
\label{eq:conv_req_amgm}
\end{align}
where (\ref{eq:conv_req_amgm}) follows from the AM-GM inequality and the fact that for a given coefficient $\theta^k(1-\theta)^{n-k}$ and variable $x_i$, there are ${n-1 \choose k-1}$ cross terms that involve $x_i$, and for a given variable $y_i$, there are ${n-1 \choose n-k-1} = {n-1 \choose k}$ cross terms that involve it. Next, using our knowledge of $\vect{x}$ and $\vect{y}$ both belonging to $S$, we obtain
\begin{align*}
\prod\limits_{i=1}^n (\theta x_i+(1-\theta y_i)) &> \theta^n + (1-\theta)^n  +
\sum\limits_{k=1}^{n-1}{n\choose k}\theta^k(1-\theta)^{n-k}\\
&=(\theta+(1-\theta))^n = 1,
\end{align*}
proving (\ref{eq:conv_req}).

To complete the proof that $f(\vect{x})$ is convex, we show that its Hessian is positive definite. The first- and second-order partials of $f$ are as follows:
\begin{align}
\frac{\partial f}{\partial x_j} &= -\frac{\prod\limits_{i\neq j}x_i}{\prod\limits_i x_i -1},\\
\frac{\partial^2 f}{\partial x_j^2} &= \left(\frac{\prod\limits_{i\neq j}x_i}{\prod\limits_i x_i -1}\right)^2,
\quad\text{and for }j\neq k,\\
\frac{\partial^2 f}{\partial x_j\partial x_k} &= \frac{-\prod\limits_{i\neq j,k}x_i\left(\prod\limits_i x_i -1\right)+\prod\limits_{i\neq j}x_i\prod\limits_{i\neq k}x_i}{\left(\prod\limits_i x_i -1\right)^2}\\
&= \frac{-\prod\limits_{i\neq j,k}x_i\prod\limits_i x_i +\prod\limits_{i\neq j,k}x_i+\prod\limits_{i\neq j}x_i\prod\limits_{i\neq k}x_i}{\left(\prod\limits_i x_i -1\right)^2}\\
&= \frac{\prod\limits_{i\neq j,k}x_i}{\left(\prod\limits_i x_i -1\right)^2}.
\end{align}
To simplify notation, we introduce
\begin{align*}
A = \prod\limits_i x_i -1,\quad B_j = \prod\limits_{i\neq j}x_i,\quad \text{and} \quad B_{j,k} = \prod\limits_{i\neq j,k}x_i,
\end{align*}
so that the second-order partials are $\partial^2 f/\partial x_j^2 = B_j^2/A^2$ and $\partial^2 f/\partial x_j\partial x_k = B_{j,k}/A^2$. 
If $\vect{x}$ has dimension $n$, the Hessian of $f(\vect{x})$ is given by
\begin{align}
H = \frac{1}{A^2}\begin{bmatrix}
B_1^2 & B_{1,2} & \cdots & B_{1,n}\\
B_{1,2} & B_2^2 & \cdots & B_{2,n}\\
\vdots & & \ddots & \vdots\\
B_{1,n} & \cdots & B_{n-1,n} & B_n^2
\end{bmatrix}.
\end{align}
For positive definiteness of $H$, we require that for any non-zero vector $\vect{y}$ of dimension $n$, $\vect{y}^TH\vect{y}>0$. \textit{I.e.}, it must be that
\begin{align*}
\frac{\vect{y}^T}{A^2}\begin{bmatrix}
y_1B_1^2 + \sum\limits_{i\neq 1} y_iB_{1,i}\\
y_2B_2^2 +\sum\limits_{i\neq 2}y_i B_{2,i}\\
\vdots\\
y_nB_n^2 + \sum\limits_{i\neq n}y_iB_{n,i}
\end{bmatrix} > 0.
\end{align*}
Since $A$ and therefore $A^2$ are strictly positive, we end up with
\begin{align}
y_1^2B_1^2 + y_1\sum\limits_{i\neq 1} y_iB_{1,i} + y_2^2B_2^2+ &y_2\sum\limits_{i\neq 2}y_i B_{2,i} + \cdots\\
+y_n^2B_n^2+y_n\sum\limits_{i\neq n}y_iB_{n,i}&>0,\nonumber\\
\sum\limits_{i=1}^n\left( y_i^2B_i^2 + y_i\sum\limits_{j\neq i}y_jB_{i,j}\right) &>0.\label{eq:ineq1}
\end{align}
Here, it is useful to introduce $P\equiv \prod\limits_{i}x_i$, and note that $B_i = P/x_i$ and $B_{i,j}=P/(x_ix_j)$. Thus, (\ref{eq:ineq1}) becomes
\begin{align}
\sum\limits_{i=1}^n\left( y_i^2\frac{P^2}{x_i^2} + \sum\limits_{j\neq i}y_iy_j\frac{P}{x_ix_j}\right) &>0.\label{eq:ineq2}
\end{align}
Letting $\gamma_i = y_i/x_i$ and dividing both sides of (\ref{eq:ineq2}) by the strictly positive $P$, we obtain
\begin{align}
\sum\limits_{i=1}^n\left( P\gamma_i^2 + \sum\limits_{j\neq i}\gamma_i\gamma_j\right) &>0.
\end{align}
Recalling that $P>1$, we note that $P\gamma_i^2 > \gamma_i^2$, $\forall i$. Thus, it would suffice to show that
\begin{align}
\sum\limits_{i=1}^n\left( \gamma_i^2 + \sum\limits_{j\neq i}\gamma_i\gamma_j\right) &>0,\\
\sum\limits_{i=1}^n\sum\limits_{j=1}^n\gamma_i\gamma_j &>0,\\
\left(\sum\limits_{i=1}^n\gamma_i\right)^2 &>0,
\end{align}
which clearly holds.\qed
\section{Non-concavity of Distillable Entanglement- and Secret Key-Based Utility Functions}
\label{app:distskfNonConc}
To show that utility functions (\ref{eq:U_D}) and (\ref{eq:U_S}) are not in general concave, it suffices to prove this for the simplest example of a single link, resulting in utilities of two variables: a rate $R$ and a Werner parameter $w$. We then must determine whether the functions ${f(R,w)\coloneqq -U^D(R,w)}$ and ${g(R,w)\coloneqq-U^S(R,w)}$ are convex.
The second-order partials of $f$ and $g$ with respect to $w$, are
\begin{align}
\frac{\partial^2f}{\partial w^2} &= \left(\frac{3}{4D_H(F)}\log_2\left(\frac{3w+1}{1-w}\right)\right)^2\nonumber\\
&\qquad-\frac{3}{D_H(F)\log(2)(3w+1)(1-w)},\label{eq:dist_sopartial}\\
\frac{\partial^2g}{\partial w^2} &= \left(\frac{1}{S_{\text{BB84}}(w)}\log_2\left(\frac{1+w}{1-w}\right)\right)^2\nonumber\\
&\qquad-\frac{2}{S_{\text{BB84}}(w)\log(2)(1+w)(1-w)}\label{eq:skf_sopartial},
\end{align}
where $F$ is given in (\ref{eq:w_to_fid}), $D_H$ is defined in (\ref{eq:DH}), and $S_{\text{BB84}}$ is defined in (\ref{eq:S_BB84}). These partials evaluate to negative values for values of $w>0.97$ (equivalently, fidelities above $0.98$ -- well within the acceptable domain of $U^D$ and $U^S$): for instance, (\ref{eq:dist_sopartial}) yields $-3.19$
and (\ref{eq:skf_sopartial}) yields $-2.34$ when $w=0.97$. Since these partials are the leading principal minors of the Hessians of $f$ and $g$, we conclude that these functions are not convex.\qed
\section*{Acknowledgment}
This work was supported in part by the NWO ZK QSC Ada Lovelace Fellowship. G.V. thanks Subhransu Maji for useful discussions on optimization.
\bibliographystyle{IEEEtran}
\bibliography{refs}
\EOD
\end{document}